\documentclass[prl,aps,epsfig,superscriptaddress,twocolumn]{revtex4}

\usepackage{bm}
\usepackage{times}
\usepackage[dvips]{graphicx}
\usepackage{mathrsfs}
\usepackage[intlimits]{amsmath}
\usepackage[colorlinks, citecolor=blue]{hyperref}
\usepackage[switch]{lineno}
\usepackage{algorithm}
\usepackage{algpseudocode}
\usepackage{setspace}
\usepackage{xcolor}
\usepackage{color}
\usepackage[defaultcolr=black]{changes}
\usepackage{ulem}
\usepackage{tabularx}

\hypersetup{linkcolor=magenta,urlcolor=blue,citecolor=blue,pdfstartview={FitH},urlcolor=blue}

\newcommand{\ket}[1]{\vert #1 \rangle}
\definechangesauthor[name={Author1}, color=red]{A2}

\begin{document}

\title{Realization of a crosstalk-free two-ion node for long-distance quantum networking}

\author{P.-C. Lai}
\altaffiliation{These authors contributed equally to this work}
\affiliation{Center for Quantum Information, Institute for Interdisciplinary Information Sciences, Tsinghua University, Beijing 100084, PR China}

\author{Y. Wang}
\altaffiliation{These authors contributed equally to this work}
\affiliation{Center for Quantum Information, Institute for Interdisciplinary Information Sciences, Tsinghua University, Beijing 100084, PR China}

\author{J.-X. Shi}
\altaffiliation{These authors contributed equally to this work}
\affiliation{Center for Quantum Information, Institute for Interdisciplinary Information Sciences, Tsinghua University, Beijing 100084, PR China}

\author{Z.-B. Cui}

\affiliation{Center for Quantum Information, Institute for Interdisciplinary Information Sciences, Tsinghua University, Beijing 100084, PR China}

\author{Z.-Q. Wang}

\affiliation{Center for Quantum Information, Institute for Interdisciplinary Information Sciences, Tsinghua University, Beijing 100084, PR China}

\author{S. Zhang}
\affiliation{Center for Quantum Information, Institute for Interdisciplinary Information Sciences, Tsinghua University, Beijing 100084, PR China}

\author{P.-Y. Liu}
\affiliation{Center for Quantum Information, Institute for Interdisciplinary Information Sciences, Tsinghua University, Beijing 100084, PR China}

\author{Z.-C. Tian}
\affiliation{Center for Quantum Information, Institute for Interdisciplinary Information Sciences, Tsinghua University, Beijing 100084, PR China}

\author{Y.-D. Sun}
\affiliation{Center for Quantum Information, Institute for Interdisciplinary Information Sciences, Tsinghua University, Beijing 100084, PR China}

\author{X.-Y. Chang}
\affiliation{Center for Quantum Information, Institute for Interdisciplinary Information Sciences, Tsinghua University, Beijing 100084, PR China}

\author{B.-X. Qi}
\affiliation{Center for Quantum Information, Institute for Interdisciplinary Information Sciences, Tsinghua University, Beijing 100084, PR China}

\author{Y.-Y. Huang}
\affiliation{Center for Quantum Information, Institute for Interdisciplinary Information Sciences, Tsinghua University, Beijing 100084, PR China}

\author{Z.-C. Zhou}
\affiliation{Center for Quantum Information, Institute for Interdisciplinary Information Sciences, Tsinghua University, Beijing 100084, PR China}
\affiliation{Hefei National Laboratory, Hefei 230088, PR China}

\author{Y.-K. Wu}
\affiliation{Center for Quantum Information, Institute for Interdisciplinary Information Sciences, Tsinghua University, Beijing 100084, PR China}
\affiliation{Hefei National Laboratory, Hefei 230088, PR China}

\author{Y. Xu}
\affiliation{Center for Quantum Information, Institute for Interdisciplinary Information Sciences, Tsinghua University, Beijing 100084, PR China}
\affiliation{Hefei National Laboratory, Hefei 230088, PR China}

\author{Y.-F. Pu}
\email{puyf@tsinghua.edu.cn}
\affiliation{Center for Quantum Information, Institute for Interdisciplinary Information Sciences, Tsinghua University, Beijing 100084, PR China}
\affiliation{Hefei National Laboratory, Hefei 230088, PR China}

\author{L.-M. Duan}
\email{lmduan@tsinghua.edu.cn}
\affiliation{Center for Quantum Information, Institute for Interdisciplinary Information Sciences, Tsinghua University, Beijing 100084, PR China}
\affiliation{Hefei National Laboratory, Hefei 230088, PR China}

\begin{abstract}
 Trapped atomic ions constitute one of the leading physical platforms for building the quantum repeater nodes to realize large-scale quantum networks. In a long-distance trapped-ion quantum network, it is essential to have crosstalk-free dual-type qubits: one type, called the communication qubit, to establish an entangling interface with telecom photons; and the other type, called the memory qubit, to store quantum information immune from photon scattering under entangling attempts. Here, we report the first experimental implementation of a telecom-compatible and crosstalk-free quantum network node based on two trapped $^{40}$Ca$^{+}$ ions. The memory qubit is encoded on a long-lived metastable level to avoid crosstalk with the communication qubit encoded in another subspace of the same ion species, and a quantum wavelength conversion module is employed to generate heralded ion-photon entanglement over a $12\,$km fiber. Our work therefore constitutes an important step towards the realization of quantum repeaters and long-distance quantum networks.
\end{abstract}

\maketitle

A future large-scale quantum network can link different functional nodes together via photonic channels \cite{kimble, weiner, ACM, BDCZ, DLCZ, rmp_gisin, rmp_polzik}, and enable numerous applications such as distributed quantum computing \cite{distributed, oxford distributed}, networked quantum sensing \cite{oxford_clock,ye_and_lukin,repeater_telescope}, and global quantum communication \cite{oxford_qkd,weifurner_qkd,BDCZ, DLCZ, rmp_gisin} with unprecedented performances. Recent advances in quantum networks have been achieved in various physical systems \cite{oxford_clock,oxford_qkd, weifurner_qkd, oxford sr, monroe_barium,oxford_sr_and_ca, 230m,lukin_siv,hanson_distillation,rempe,2021nature,bao3nodes,lukin reflection}. Among these different physical platforms, trapped atomic ions constitute one of the most promising systems to build quantum repeater nodes for scalable quantum networks, as the highest fidelities in quantum logic operations and state detection \cite{oxford_gate, quantiuum} have been demonstrated on this platform, which are important for the entanglement swapping \cite{lanyon_connection,Eschner,lattice}, distillation \cite{lukin_siv,hanson_distillation}, and error correction \cite{quantiuum} required in a large-scale quantum network. Other advantages include the long coherence time at the scale of an hour \cite{1hour}, and the record-high fidelity and rate in entangling two network nodes \cite{oxford sr, monroe_barium}.

To enable a long-distance trapped-ion quantum network, crosstalk-free dual-type qubits on each node and the interface of the node with the telecom photons are two essential requirements \cite{duan pra,duan arxiv, duan rmp}. For dual-type qubits on each node, it is required that the memory qubits are employed to store quantum information for later computational tasks or establishment of remote entanglement, while the communication qubits are interfaced with photons to link different network nodes or used occasionally for sympathetic cooling. It is noteworthy that the most critical crosstalk in a trapped-ion quantum network is from the spontaneously emitted photon during the cooling, pumping, detection and ion-photon entangling on the communication qubit. These scattered photons can be reabsorbed by the memory qubit, and the stored quantum information is ruined. Thus this crosstalk can only be solved by spectral isolation, and cannot be solved even with perfect individual addressing. The memory and the communication qubits can be implemented either by employing two different ion isotopes \cite{oxford_sr_and_ca, monroe_barium}, or by the same ion species but exploiting a dual-type encoding with two sets of spectrally separated qubit states \cite{yanghx,huangyy,omg}. Another requisite for a long-distance quantum network is that the network node should be interfaced with photons in the telecom band to minimize the transmission loss in optical fibers. This requirement has been fulfilled either by employing atomic transitions at the telecom band \cite{erbium, thompson erbium, kuzmich telecom,changwei}, or by converting the emitted photon to the telecom wavelength on an additional module \cite{lanyon_connection,weifurner_qkd,keller,ikuta,2021nature, Eschner long}. Implementation of the above two essential requirements for a long-distance quantum network together in a single node is experimentally very challenging and has not been achieved so far in a trapped-ion system.

In this work, we report the first experimental realization of a crosstalk-free and telecom-compatible quantum network node based on two $^{40}$Ca$^{+}$ ions in a dual-type style. The memory qubit is encoded on a metastable level which is unaffected by the photon-scattering operations on the communication qubit due to the large frequency difference. The two types of qubits can be coherently converted back and forth by a Raman and a quadrupole transition. A quantum frequency conversion module is employed to convert the near-infrared photon emitted from the ion to the telecom band, and the heralded ion-photon entanglement has been demonstrated over different fiber lengths at $3\,$m, $1\,$km, and $12\,$km. We characterize the performance of the memory qubit with the measured coherence time of about $323(19)\,$ms and confirm that quantum information is protected in the memory qubit under dissipative operations on the communication qubit. With all these achievements, we implement a promising quantum node based on trapped ions which fulfils the essential requirements for a long-distance quantum network.

\begin{figure}
  \centering
  \includegraphics[width=8.7cm]{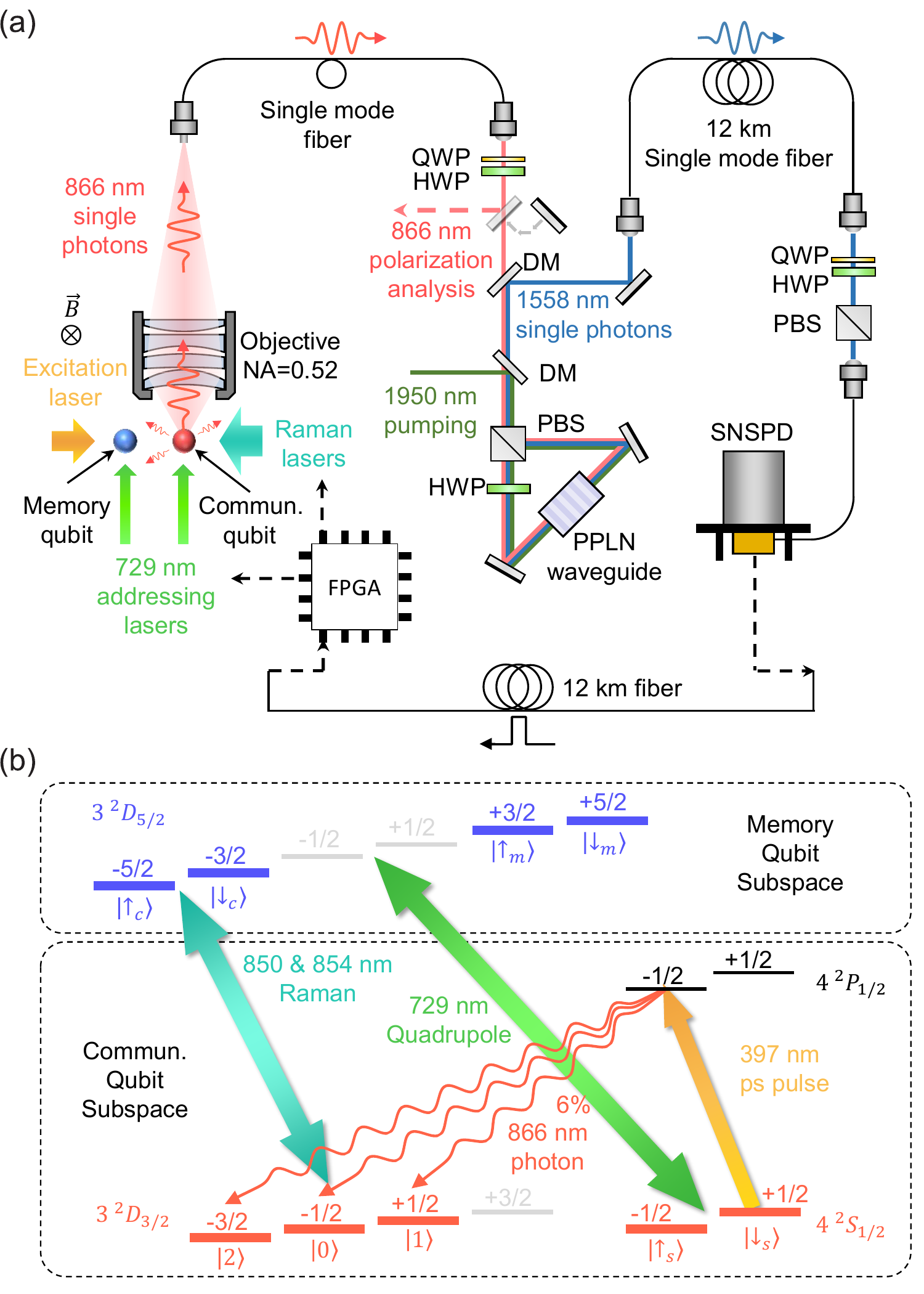}\\
  \caption{Experimental setup and crosstalk-free qubit subspaces.
  (a) The two-ion quantum network node consists of a communication qubit and a memory qubit encoded in different subspaces of $^{40}$Ca$^+$ ion. The $729\,$nm beam is individually addressed. All other beams are global. (b) Two spectrally-separated subspaces for the communication qubit and the memory qubit. The communication qubit subspace is a closed three-level cycle, and the memory qubit subspace is a metastable level. Bridges connecting these two subspaces include $850\,\text{nm}/854\,\text{nm}$ Raman and $729\,\text{nm}$ quadruple transitions. Ion-photon entanglement is excited by a $397\,\text{nm}$ picosecond pulse.
  }
\end{figure}

\begin{figure}
  \centering
  \includegraphics[width=8.7cm]{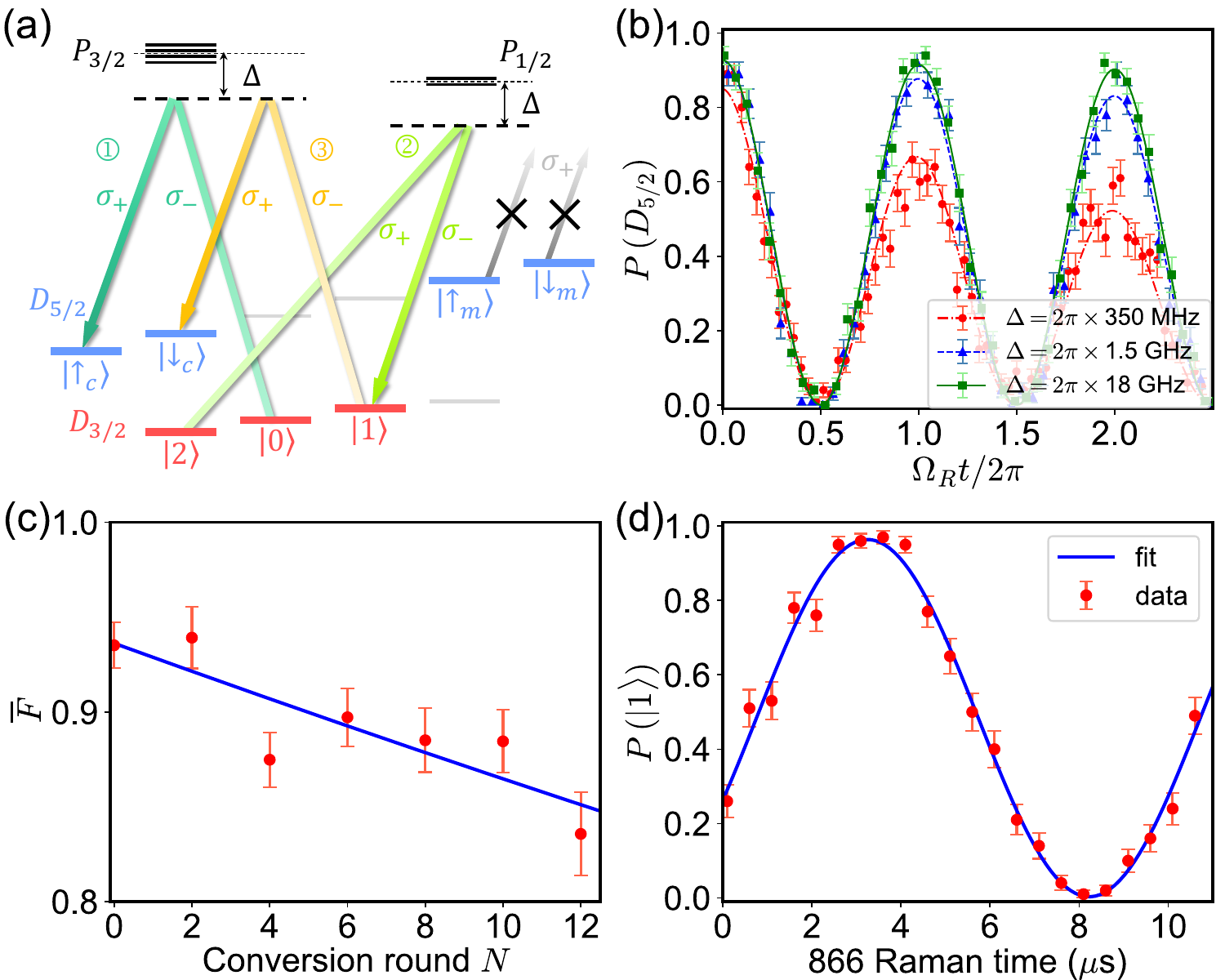}\\
  \caption{Connecting communication and memory qubit subspaces through Raman transitions.
  (a) After a photon is detected, we use three successive Raman transitions to combine two energy levels related to $\ket{\sigma^{\pm}}$ photon emissions and map the ion state from the communication qubit subspace to the memory qubit subspace for further state manipulation and detection. (b) We measure the probability of unwanted excitation to the upper state during the Raman process with variable detunings at $350\,$MHz, $1.5\,$GHz and $18\,$GHz. We prepare the ion to $|\uparrow_c\rangle$ and perform a Raman pulse driving the transition $|\uparrow_c\rangle\leftrightarrow|0\rangle$ for different time and detuning. The chance of off-resonant excitation to the upper state is $12(1)\%$, $2.7(3)\%$ and $0.7(2)\%$ for detunings at $350\,$MHz, $1.5\,$GHz and $18\,$GHz every $\pi$-pulse. (c) We measure the state transfer fidelity between the memory qubit subspace bases $\ket{\uparrow_c}/\ket{\downarrow_c}$ and the communication qubit subspace bases $\ket{0}/\ket{1}$, with different rounds of $\pi$ pulses. The fitted $\pi$-pulse fidelity is $99.2(2)\%$ averaged on six mutually unbiased states. (d) Rabi oscillation driven by an$866\,\text{nm}/866\,\text{nm}$ Raman transition. The initial $\frac{1}{2}|1\rangle+\frac{\sqrt{3}}{2}|2\rangle$ state can be combined to $\ket{1}$ with a fidelity of $96(2)\%$ by a $\frac{2}{3}\pi$-pulse.
  }
\end{figure}

The quantum network node is based on two $^{40}$Ca$^{+}$ ions in a segmented blade trap, as shown in Fig.~1(a). Here the communication qubit stays in a closed $3$-level subspace containing $|S_{1/2}\rangle$, $|D_{3/2}\rangle$, and $|P_{1/2}\rangle$ (see Fig.~1(b)), so that all the dissipative operations on the communication qubit including cooling, optical pumping, fluorescence detection and photon excitation can be carried out by lasers at $397\,$nm and $866\,$nm in this subspace, and the fluorescent photons scattered from the communication qubit are also at these two wavelengths (or at $732\,$nm with a very rare chance, see Fig.~S1 for the full energy levels and transitions). On the other hand, the metastable level $|D_{5/2}\rangle$ can only be driven by two transitions at $854\,$nm or $729\,$nm, which are perfectly isolated in frequency from the lasers or scattered photons for the operations on the communication qubit. Therefore we use $|D_{5/2}\rangle$ subspace to carry the memory qubit.

\begin{figure*}
  \centering
  \includegraphics[width=17.6cm]{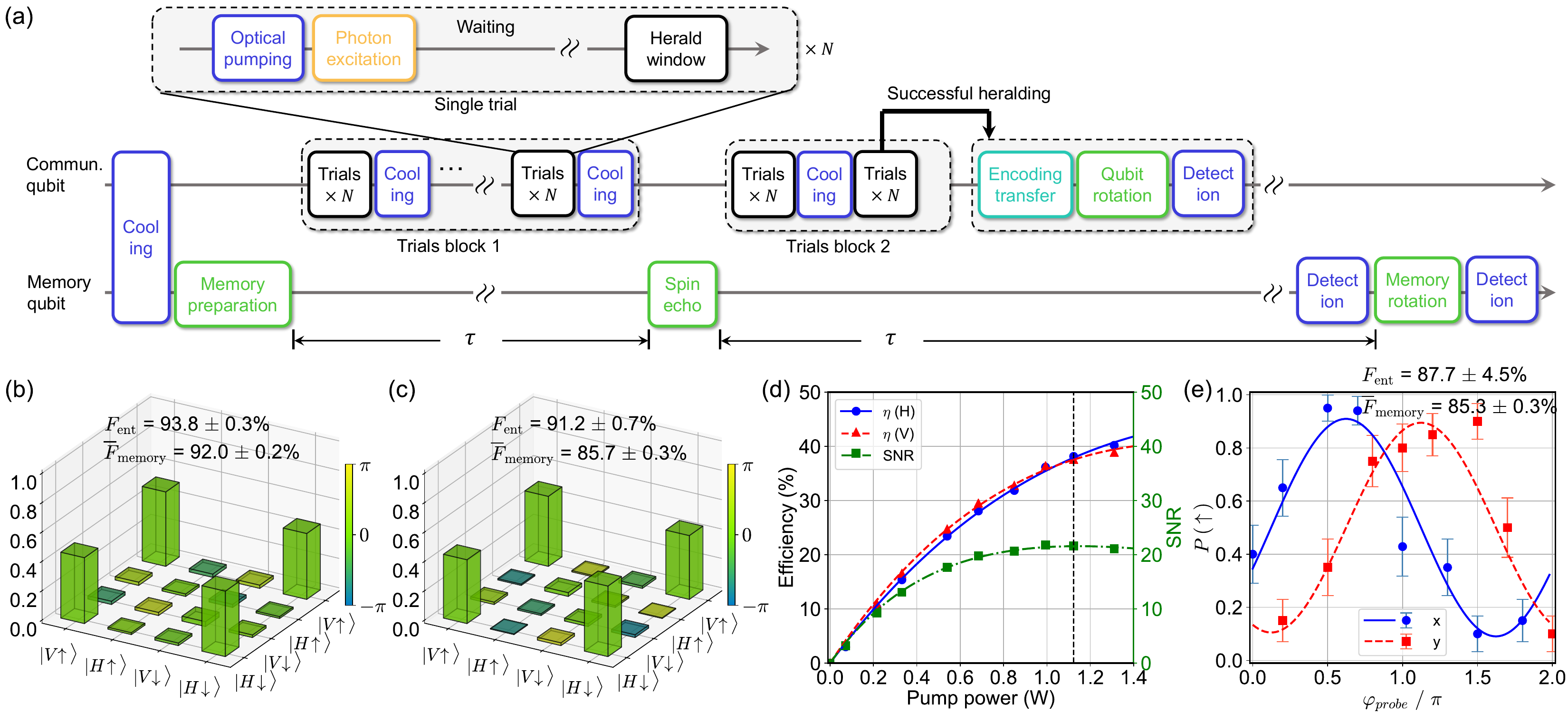}\\
  \caption{The experimental protocol and heralded ion-photon entanglement at different fiber lengths.
  (a) The protocol used in this experiment. In each sequence, both ions are measured. (b-c) Reconstructed density matrix for the ion-photon entangled state with fiber lengths of $3\,\text{m}/1\,\text{km}$ and $2\tau=40\,\text{ms}/200\,\text{ms}$. The ion-photon entanglement fidelity (regarding communication qubit) and the memory qubit fidelity (averaged over six MUBs) are $93.8\pm0.3\%$ and $92.0\pm0.2\%$ in the $3\,$m case ($91.2\pm0.7\%$ and $85.7\pm0.3\%$ for the $1\,$km case) respectively, as shown in Fig.~3(b, c). (d) The performance of frequency conversion. We use laser power at about $1.1\,$Watts (dashed line) for each polarization. (e) Measure the coherence of the ion-photon entangled state over $12\,$km fiber by projecting photons at $\ket{H}+\ket{V}$ (blue circle) or $\ket{H}+i\ket{V}$ bases (red square) and measuring the visibility of sinusoidal oscillations with different phases of $729\,\text{nm}$ rotations. The ion-photon entanglement fidelity (regarding communication qubit) and the memory qubit fidelity (averaged over six MUBs with a storage time of $200\,$ms) are $87.7\pm4.5\%$ and $85.3\pm0.3\%$ in the $12\,$km case, as shown in Fig.~3(e).
  }
\end{figure*}

There are two methods to connect these two qubit subspaces, as shown in Fig.~1(b). (Note that in the experiments demonstrated later in this paper, both methods are used for the ion-photon entanglement generation and state characterization.) The first is the standard quadrupole transition at $729\,$nm which connects the $|S_{1/2}\rangle$ level in the communication qubit subspace and $|D_{5/2}\rangle$ level in the memory qubit subspace. The $729\,$nm beam is focused on each ion with a Gaussian radius of $2\,\mu$m via an addressing system based on two crossed acoustic optical deflectors (AODs) and another objective. The typical fidelity of a $729\,$nm $\pi$-pulse is about $99\%$ in this experiment.

The second bridge is a Raman transition by two lasers at $850\,$nm and $854\,$nm, which connects $|D_{3/2}\rangle$ level and $|D_{5/2}\rangle$ level in different subspaces, as shown in Fig.~2(a). The two beams co-propagate along the magnetic field axis in complementary circular polarizations, which drive Raman transitions between $|D_{3/2},m\rangle$ and $|D_{5/2},m-2\rangle$ ($m=-1/2,+1/2,+3/2$). Here we set the Raman detuning $\Delta$ to $18\,$GHz with the chance of unwanted excitation to the intermediate state $\ket{P_{3/2}}$ at about $0.7(2)\%$ per $\pi$-pulse, as illustrated in Fig.~2(b). We demonstrate the coherent state transfer between a memory qubit subspace encoding on $|\uparrow_c\rangle/|\downarrow_c\rangle\equiv|D_{5/2},m=-5/2\rangle/|D_{5/2},m=-3/2\rangle$ and a communication qubit subspace encoding on $|0\rangle/|1\rangle\equiv|D_{3/2},m=-1/2\rangle/|D_{3/2},m=+1/2\rangle$, by two successive Raman $\pi$-pulses which drive the transitions $|\uparrow_c\rangle\leftrightarrow|0\rangle$ and $|\downarrow_c\rangle\leftrightarrow|1\rangle$ respectively. We measure the average transfer fidelity by initializing the qubit to six mutually unbiased states (MUB) $\ket{\uparrow}$, $\ket{\downarrow}$, $\ket{+}$, $\ket{-}$, $\ket{L}$, $\ket{R}$ encoded on $|\uparrow_c\rangle/|\downarrow_c\rangle$, and applying variable numbers of Raman $\pi$-pulses. The average transfer fidelity is measured to be $99.2(2)\%$ per Raman $\pi$-pulse for six MUBs (mutually unbiased bases) (see Fig.~2(c)).

The network protocol is illustrated in Fig.~3(a). (For all the results demonstrated below, both the ion-photon entanglement fidelity regarding communication qubit and the memory qubit fidelity are measured in a single experiment following the sequence as shown in Fig.~3(a).) After $2.5\,$ms Doppler cooling and $1.4\,$ms EIT cooling, we prepare the memory qubit to a superposition of $\ket{\uparrow_m}=\ket{D_{5/2},m=+3/2}$ and $\ket{\downarrow_m}=\ket{D_{5/2},m=+5/2}$ via optical pumping and $729\,$nm pulses. The memory qubit is stored for two periods of $\tau$ with a spin echo in the middle (Section S10), and read out at the end. Inside each period of $\tau$, repetitive trials of ion-photon entangling attempts are performed in a heralded style via an excitation pulse on the communication qubit. Once a photon detection event is recorded, the excitation trial halts immediately and state inspection is performed on the communication qubit via the Raman conversion, state manipulation, and fluorescence detection. The intermediate cooling is performed every $N=100$ trials. In each trial, the communication ion is first prepared to the initial state $\ket{S_{1/2},m=+1/2}$ by $290\,$ns of optical pumping. A strong picosecond pulse at $397\,$nm pumps the ion to the excited state $\ket{P_{1/2},m=-1/2}$ with nearly unity efficiency. The excited state then decays to $\ket{D_{3/2}}$ level with a branching ratio of $6\%$ and emits a photon at $866\,$nm (see Fig.~1(b)) \cite{branching ratio}. Once this happens, the ion and the photon are prepared into an entangled state, which can be expressed as:
\begin{equation}
|\Psi\rangle =\frac{1}{\sqrt{3}}|0\rangle|\pi\rangle+\frac{1}{\sqrt{6}}|1\rangle|\sigma^-\rangle+\frac{1}{\sqrt{2}}|2\rangle|\sigma^+\rangle
\end{equation}
where $|0\rangle\equiv|D_{3/2},m=-1/2\rangle\rangle$, $|1\rangle\equiv|D_{3/2},m=+1/2\rangle$, $|2\rangle\equiv|D_{3/2},m=-3/2\rangle$, and the weights are from the Clebsh-Gordan coefficients of the corresponding transitions. The emitted photon is collected by an objective of NA $=0.52$ with the optical axis perpendicular to the magnetic field, and then coupled into a single-mode fiber. In this case, the emission in the $\ket{\pi}$ polarization will be projected to $\ket{H}$, and $\ket{\sigma^\pm}$ will be projected to $\ket{V}$ with a half probability. Thus the entangled state becomes $|\Psi\rangle =\frac{1}{\sqrt{2}}[|0\rangle|H\rangle+(\frac{1}{2}|1\rangle+\frac{\sqrt{3}}{2}|2\rangle)|V\rangle]$. We further combine $\frac{1}{2}|1\rangle+\frac{\sqrt{3}}{2}|2\rangle$ into $\ket{1}$ and convert the ion state to the memory qubit subspace via three successive Raman transitions, including ($1$) an $850\,\text{nm}/854\,\text{nm}$ Raman converting $\ket{0}$ to $\ket{\uparrow_c}$, ($2$) an $866\,\text{nm}/866\,\text{nm}$ Raman (Section S9) which converts $\frac{1}{2}|1\rangle+\frac{\sqrt{3}}{2}|2\rangle$ to $\ket{1}$ (see Fig.~2(d)), and ($3$) another $850\,\text{nm}/854\,\text{nm}$ Raman converting $\ket{1}$ to $\ket{\downarrow_c}$ (see Fig.~2(a)). After these Raman transitions, we end up with a maximally entangled Bell state:
\begin{equation}
|\Psi\rangle =\frac{1}{\sqrt{2}}(|\uparrow_c\rangle|H\rangle+|\downarrow_c\rangle|V\rangle)
\end{equation}
The reason for transferring the communication ion to the memory qubit subspace is that we can use standard $729\,\text{nm}$ rotations for the qubit state inspection in the next step. If needed, it is not hard to transfer the entangled state back to communication qubit subspace $|\Psi\rangle =\frac{1}{\sqrt{2}}(|0\rangle|H\rangle+|1\rangle|V\rangle)$ via additional $850\,\text{nm}/854\,\text{nm}$ Raman pulses, or to $|\Psi\rangle =\frac{1}{\sqrt{2}}(|\uparrow_s\rangle|H\rangle+|\downarrow_s\rangle|V\rangle)$ via $729\,\text{nm}$ pulses ($|\uparrow_s\rangle/|\downarrow_s\rangle\equiv\ket{S_{\frac{1}{2}},m_s=\mp\frac{1}{2}}$).

\begin{figure}
  \centering
  \includegraphics[width=8.7cm]{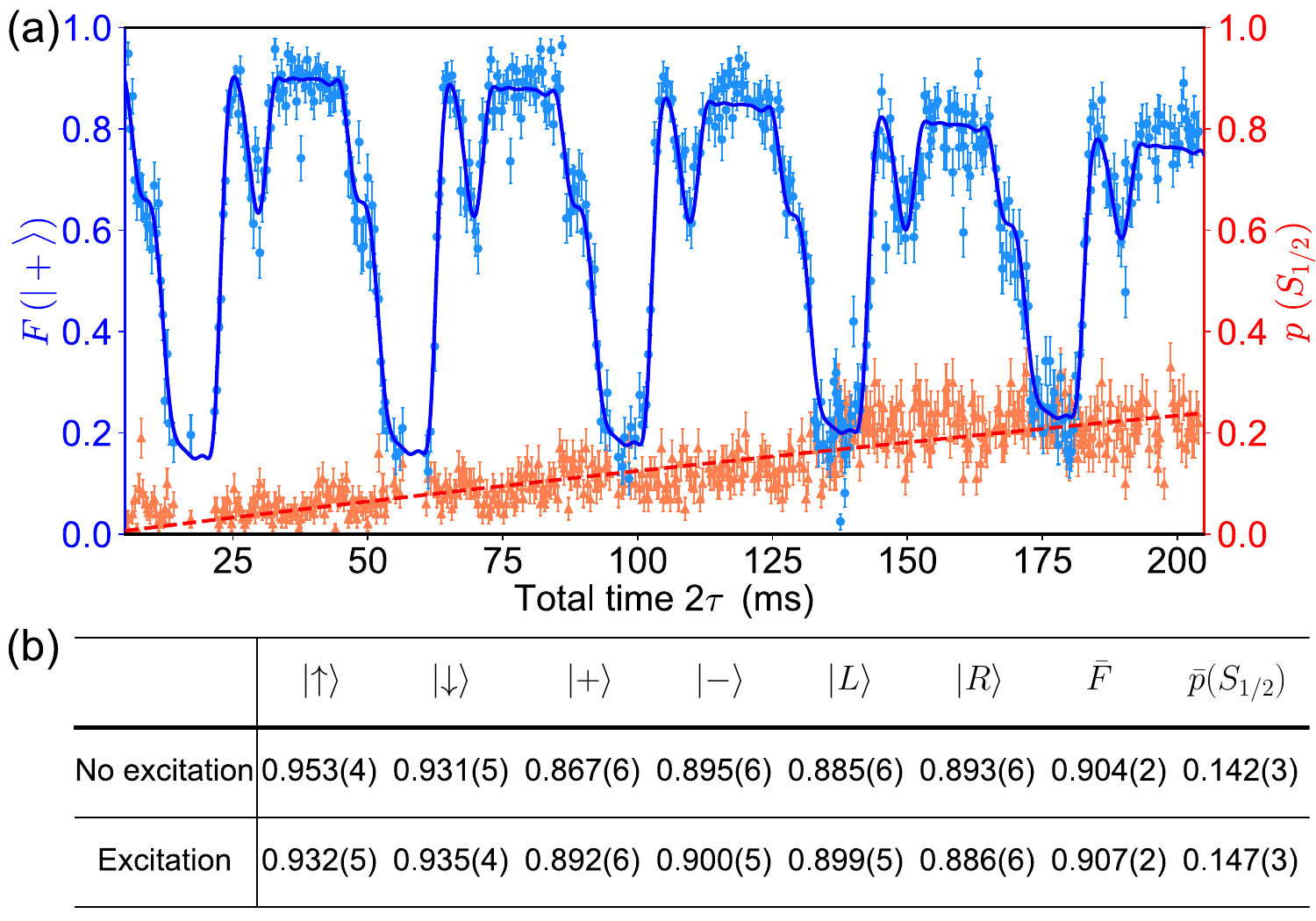}\\
  \caption{Crosstalk-free storage of quantum information in memory qubit.
  (a) We measure the decay probability (orange triangle) and the storage fidelity to the initial state $\frac{\ket{\uparrow_m}+\ket{\downarrow_m}}{\sqrt{2}}$ (blue circle) with variable storage time, in the protocol shown in Fig.~3(a). (b) The fidelities for six MUBs, average fidelity, and decay probability with/without repetitive cooling, pumping, photon excitation, and state detection on the communication qubit. Note that the unchanged storage fidelity is also an evidence that the intermediate cooling on communication qubit has efficiently cooled the motional state of the memory ion. Otherwise the heated motional state will significantly influence the fidelity of the rotation pulses and the final fidelity of the stored state.
  }
\end{figure}

The emitted photon at $866\,\text{nm}$ can be further converted to $1558\,\text{nm}$ on a PPLN (Periodically Poled Lithium Niobate) waveguide surrounded by a Sagnac interferometer in polarization-maintained style~\cite{weifurner_qkd,ikuta}. The external device efficiency at $38\%$ and a signal-to-noise-ratio (SNR) of $22$ are achieved by $2.2\,$W of pumping laser, as shown in Fig.~3(d).

We measure the ion-photon entanglement fidelity for three different fiber lengths, i.e. $866\,\text{nm}$ photon over a $3\,$m fiber, $866\,\text{nm}$ photon over a $1\,$km fiber, and converted $1558\,\text{nm}$ photon over a $12\,$km fiber. We use quantum state tomography to characterize the quality of the ion-photon entanglement at $3\,$m and $1\,$km. The density matrices of the ion-photon state in these two cases are reconstructed via the MLE (maximum-likelihood estimate) and are illustrated in Fig.~3(b) and 3(c). The measured fidelity of ion-photon entanglement (regarding communication qubit) to the most close maximally-entangled state is $93.8(3)\%$ over a $3\,$m fiber and the memory qubit has an averaged fidelity of $92.0\pm0.2\%$ over $6$ MUBs with a total storage time $2\tau=40\,$ms in memory qubit. The fidelity decays a little bit to $92.4(4)\%$ at longer storage time $2\tau=200\,$ms, mainly due to increased instability of $729\,\text{nm}$ laser power (with memory qubit fidelity decays to $86.4\pm0.3\%$). At a fiber length of $1\,$km, the ion-photon entanglement fidelity (regarding communication qubit) is measured to be $91.2(7)\%$, with a storage time of $200\,$ms in memory qubit (memory qubit fidelity at $85.7\pm0.3\%$). For the $12\,$km ion-photon entanglement with $200\,$ms storage time in memory qubit, we characterize the fidelity by measuring the visibilities for the ion-photon entangled state in three bases $\sigma_x$, $\sigma_y$ and $\sigma_z$ \cite{pan50km}. By adjusting the phase of the final $729\,\text{nm}$ $\pi/2$ rotation, we could observe two sinusoidal oscillations in the $\ket{H}+\ket{V}$ and $\ket{H}+i\ket{V}$ bases as shown in Fig.~3(e) with visibilities of $V_x=81.9\pm10.7\%$ and $V_y=79.0\pm10.4\%$. Thus the entanglement fidelity can be estimated as $F=\frac{1}{4}(1+V_x+V_y+V_z)=87.7\pm4.5\%$, with $V_z=89.8\pm10.7\%$. The memory qubit fidelity is $85.3\pm0.2\%$ in this case. The generation rate of heralded ion-photon entanglement over $3\,$m, $1\,$km, $12\,$km fiber is $46\,\text{s}^{-1}$, $3.4\,\text{s}^{-1}$, $0.032\,\text{s}^{-1}$, and the attempting rate is $264\,\text{kHz}$, $49\,\text{kHz}$, $5\,\text{kHz}$, respectively. We also discuss the potential improvements in the success rate and the schemes to generate ion-ion entanglement in the future in Section S7, S8 and S12.

In the last step, we verify that the memory ion can protect the stored quantum information from the dissipative operations on the communication qubit. Note that the pulse sequences applied to the communication qubit are identical to the pulses applied during the entanglement experiments demonstrated above.  Unlike the ground state, the metastable $\ket{D_{5/2}}$ level has a finite lifetime and will decay to $\ket{S_{1/2}}$ in a timescale of $T_1\sim1\,$s, which means the stored quantum information will suffer from noticeable degradation if the storage time is comparable to $T_1$. The good news is that this error can be probed via fluorescence detection, which means once happened, it can be discarded to avoid influencing the fidelity of the network task, but at a cost of deteriorated efficiency \cite{kauffman midcircuit, midcircuit2, midcircuit3, thompson erasure, quantiuum}. In the protocol of this experiment, after a storage time of $2\tau$, the memory qubit is analyzed first by a fluorescence detection to probe whether it has decayed to $\ket{S_{1/2}}$. If the memory ion is in the bright state, the protocol terminates and starts over. If the memory ion is in the dark state, we rotate the ion state and execute the second state detection to perform single-qubit state tomography on the quantum state.

We characterize the probability of the level decay by the first state detection with variable storage time $2\tau$, and the fitted lifetime of the memory qubit is found to be $T_1'=0.79(3)\,$s (Fig.~4(a)). We attribute the deviation from the precise lifetime $T_1=1.168(7)\,$s of $\ket{D_{5/2}}$ \cite{lifetime} to the leakage from $854\,\text{nm}$ and $729\,\text{nm}$ beams. We further initialize the memory ion to $\ket{+}=\frac{\ket{\uparrow_m}+\ket{\downarrow_m}}{\sqrt{2}}$ at the beginning of the protocol, and measure the storage fidelity with variable time $2\tau$, conditioned on that the memory qubit is still in $\ket{D_{5/2}}$ to exclude the influence from $T_1'$. This spontaneous decay causes about $4.9\%/22.4\%/22.4\%$ reduction in the success rate of the protocol (ion-photon entanglement is successfully heralded and memory qubit does not decay) for the $3\,\text{m}/1\,\text{km}/12\,\text{km}$ cases. In this way, the memory dephasing time is found to be $T_2=323(19)\,$ms, as shown in Fig.~4(a). The fidelity oscillates with the storage time, and can be well fitted by a double-frequency noise model at $50\,$Hz and $150\,$Hz (analyzed in Section S16). To confirm that the operations performed on the communication qubit do not influence the memory qubit, we measure the storage fidelities of six mutually unbiased bases (MUBs) with $2\tau=120\,$ms, with or without all the dissipative operations on the communication qubit. We find the average storage fidelity of six bases is $0.907(2)/0.904(2)$ with/without operations on the communication ion, and the averaged decay probability is $0.147(3)/0.142(3)$ with/without operations on the communication ion. Since both of the storage fidelity and decay probability cannot be distinguished considering the statistical error, we confirm the operations on the communication ion do not influence the memory ion. (Despite crosstalk depending linearly on storage time, due to negligible infidelity contribution at $120\,$ms, it is safe to assume it will also be negligible in the $40\,$ms/$200\,$ms cases. We also theoretically estimate the crosstalk influence on memory qubit's internal state and motional state in Section S13 and S14. The infidelity induced by the crosstalk is at the level of $10^{-5}$ in the $120\,$ms case.)

In summary, we experimentally realize a novel two-ion quantum network node, which can be performed in a crosstalk-free style and is suitable for distributing quantum entanglement over a metropolitan-scale fiber network. Our scheme is useful not only for $\text{Ca}^{+}$ (which has advantages in transition wavelength for conversion to telecom C band, see Section S6), but also suitable for other physical platforms which have long-lived metastable energy levels such as $\text{Ba}^{+}$ (with a favorable branching ratio of $493\,$nm transition~\cite{quraishi}), $\text{Sr}^{+}$~\cite{oxford sr} and neutral $^{171}$Yb atoms in a tweezer array \cite{kauffman midcircuit, midcircuit2, midcircuit3, thompson erasure}. In Section S17, we compare the dual-type scheme (this work) with the dual-species scheme (\cite{oxford distributed,oxford_sr_and_ca}) in terms of the crosstalk effect and the capability to support distributed quantum computing. It is also important to further improve the remote ion-photon entanglement generation rate. We believe that the enhancements through a cavity \cite{230m, lanyon_connection, keller, Takahashi, rempe, thompson erbium} or via multiplexing \cite{225,multipurpose,2021nature,rmp_gisin,enhancement, lanyon multiplexing, haffner multiplexing} are promising approaches towards this goal. Together with the techniques such as phase stabilization, deployed fiber, and the synchronization of control systems developed in recent works on metropolitan quantum network~\cite{bao3nodes,hanson long,lukin reflection,230m,Eschner long}, it is possible to build a metropolitan trapped-ion quantum network based on the network node demonstrated in this work in the future. We also demonstrate the scheme to implement memory-enhanced connection of two ion-photon entangled segments in Section S11.

We thank Lai Zhou, Chi Zhang, Lu Feng, Panyu Hou, and Ye Wang for the helpful discussions.
This work is supported by Innovation Program for Quantum Science and
Technology (No.2021ZD0301604), the Tsinghua University
Initiative Scientific Research Program and the Ministry of
Education of China through its fund to the IIIS. Y.F.P. acknowledges
support from the Dushi Program
from Tsinghua University.


\onecolumngrid

\setcounter{equation}{0}
\setcounter{figure}{0}
\setcounter{table}{0}
\setcounter{page}{1}
\setcounter{section}{0}
\makeatletter
\renewcommand{\theequation}{S\arabic{equation}}
\renewcommand{\thefigure}{S\arabic{figure}}
\renewcommand{\thetable}{S\arabic{table}}

\pagebreak
\begin{center}
\Large Supplemental Material for\\\textbf{Realization of a crosstalk-free multi-ion node for long-distance quantum networking}
\end{center}

\subsection{Section S1: Level structure and laser beam diagram}
Fig.~S1 shows the level structure of ${}^{40}\mathrm{Ca}^+$. In our experiment, the ion is initialized in $\ket{S_{1/2},m = + 1/2}$ by optical pumping with $\sigma^{+}$-polarized $397\,\text{nm}$ continuous laser beam and $866\,\text{nm}$ beam. Doppler cooling, electromagnetically induced transparency (EIT) cooling and fluorescence detection are also performed by $397\,\text{nm}$ laser beam and $866\,\text{nm}$ beam.

The trapping frequencies of our trap are $\text{(}\omega_x, \omega_y, \omega_z\text{)}=2\pi\times\text{(}1.65, 1.55, 0.5\text{)}\,$MHz in three dimensions and the separation between two ions is $9\,\mu$m. The magnetic field is set to $4.2\,$G in the experiment.

\begin{figure}[h]
  \centering
  \includegraphics[width=0.6\linewidth]{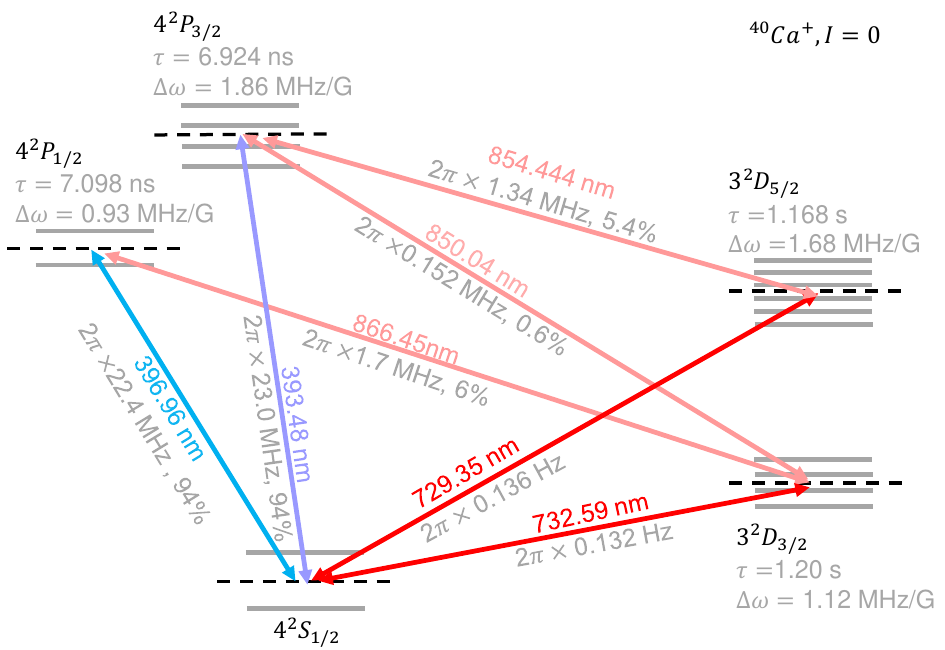}
  \caption{${}^{40}\mathrm{Ca}^+$ ion energy level structure \cite{stute thesis}. We list all the energy levels and transitions of ${}^{40}\mathrm{Ca}^+$ ion in this figure.}

\end{figure}

\subsection{Section S2: Control system}

For specific purpose of heralded entanglement experiments, a home-made device ``QNetWorker'' handling the entanglement attempts with high speed and precision in decision making process has to be developed, in cooperation with the other programmable logic unit ``Sequencer", which controls usual sequences and offers the flexibility of managing the overall branch of decision for all qubits to enable swift scheduling of entanglements generation and subsequent operation.

The QNetWorker functions in entanglement establishment alongside encoding transfer for network qubits with logic controls as the following content describes:

\begin{itemize}
    \item TTL outputs for controlling $397\,$nm and $866\,$nm laser beam's RF switches, for driving corresponding AOM for state preparation and providing the pulse picker with EOM trigger, which composes the trial sequence up to $1\,\text{MHz}$ repetition rate, mainly limited by AOM latency.
    \item Four input channels with separately adjustable time window for receiving signal of SNSPD. Facilitating timestamps readout of heralded events with superior precision of  $417\,$ps  ($2\,$ns adopted in this work).
    \item Decisions of consecutive sequences to manipulate communication qubits based on heralding patterns encoded in a $4$-bit word of input events without latency. Moreover, the subsequent sequences are synchronized with the timestamps, for minimized $866\,\text{nm}$ Raman transition phase error for merging gate.
    \item Redundant interfaces of modules facilitate the potential to programmable modulations by adequate digital output channels, such as feedforward phase correction of $866\,\text{nm}$ Raman merging gate (not in this work) and controlling sequences of arbitrary waveform generator based on heralding patterns and experiment design.
\end{itemize}

Furthermore, Sequencer serves as the master controller of experiment sequences which regulates QNetWorker for entangling trials. It enables immediate detection right after heralding events or in-sequence manipulations and detection of communication qubits during arbitrary gate sequences such as spin echo of memory qubits.  In principle, any sophisticated gates implementation could be accomplished in current system.

\begin{figure}[h]
  \centering
  \includegraphics[width=0.6\linewidth]{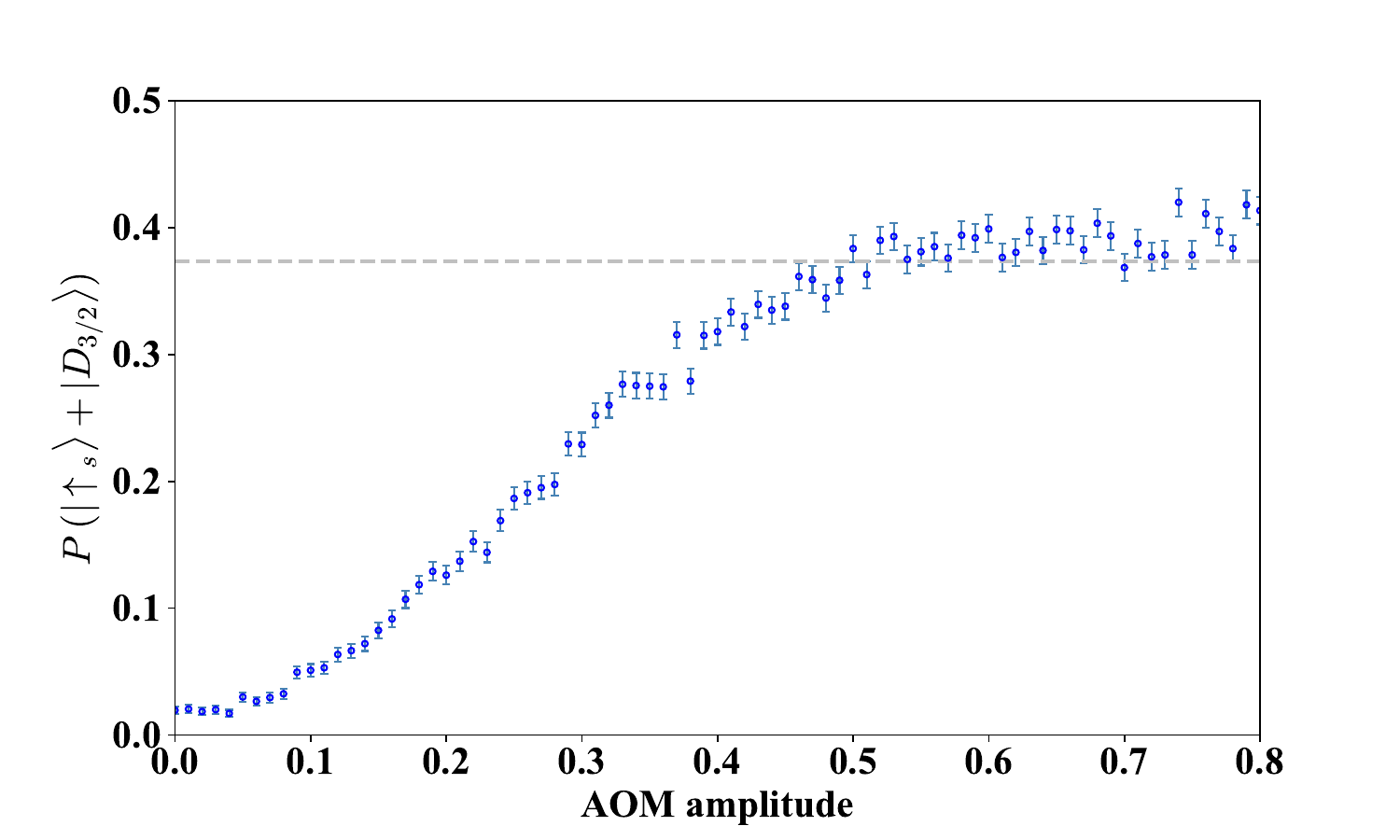}
  \caption{Calibration of excitation pulse. We verify a single $397\,$nm picosecond pulse can drive a $\pi$-pulse by measuring the proportion of state pumped to $\ket{\uparrow_s}\equiv\ket{S_{1/2},m=-1/2}$ and $\ket{D_{3/2}}$ after excitation, with variable pulse energy. Here the nonzero probability at $0$ AOM amplitude originates from  the imperfect optical pumping and state detection. Current measurement is already sufficient to find out the rough AOM amplitude corresponding to a $\pi$-pulse for the ion-photon entanglement excitation. In the future this can be improved by polishing the optical pumping and state detection.}

\end{figure}

\subsection{Section S3: Pulse excitation}
In the experiment, we use $397\,$nm pulse laser with $\sigma^{-}$ polarization to excite ion from $\ket{S_{1/2},m = +1/2}$ to $\ket{P_{1/2},m =-1/2}$. After pulse excitation and spontaneous decay of ion, population in $\ket{S_{1/2},m = -1/2}$ and $\ket{D_{3/2}}$ is measured. The outcome of the calibration process is shown in Fig.~S2, with which we get the corresponding AOM amplitude to achieve pulse area of $\pi$. The dashed line shows the theoretical population in $\ket{S_{1/2},m =-1/2}$ and $\ket{D_{3/2}}$ for $\pi$-pulse. State preparation and shelving error account for the fraction exceeding theoretical value. After determination of $\pi$-pulse AOM amplitude, we verify that the $397\,$nm pulse laser will not off-resonantly drive the ion to $\ket{P_{3/2}}$, and induce the ion to fall out of the $3$-level communication subspace, which results in $\ket{3D_{5/2}}$ population. We excite ion from $\ket{S_{1/2},m =+1/2}$ to $\ket{P_{1/2},m =-1/2}$ repeatedly and measure the state of ion. Pulse excitations for over $10^6$ times gives no $3D_{5/2}$ population, which verifies pulse laser in our experiment will not give rise to excitation from $\ket{4S_{1/2}}$ to $\ket{4P_{3/2}}$.

We plot the histogram for the arrival time of $866\,$nm photon relative to the pulse excitation trigger. Fig.~S3 shows the time distribution of photons collected. Latency, jitters and exponential decay of ion give rise to the shape of the histogram. We fit the histogram with convolution of Gaussian distribution and exponential decay, and the fitted $1/e$ decay time is $6.936\pm 0.147\,$ns for $3\,$m case, $7.091\pm 0.082\,$ns for $1\,$km case, and $6.703\pm 0.143\,$ns for $12\,$km case. We didn't observe remarkable difference in photon shape with different fiber lengths. In the experiment, we record timestamps of photons detected by single photon detector and optimize the detection window accordingly to achieve a high signal-to-noise ratio.

\begin{figure}[h]
    \centering
    \includegraphics[width=1\linewidth]{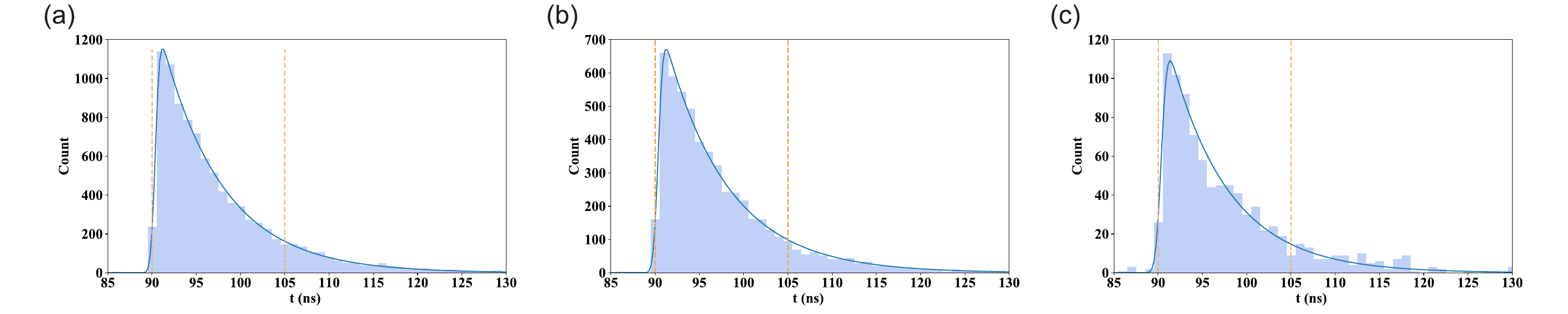}
    \caption{Photon distribution. The histogram of photon arrival time relative to the excitation trigger. The $15\,$ns acceptance window of photons is between the two dashed lines. The fitted $1/e$ decay time is $6.936\pm 0.147\,$ns for (a) $3\,$m case, $7.091\pm 0.082\,$ns for (b) $1\,$km case, and $6.703\pm 0.143\,$ns for (c) $12\,$km case.}

\end{figure}

\subsection{Section S4. 850 \& 854 Raman}
Fig.~S4(a) shows $850\,$nm and $854\,$nm laser locking scheme. They are locked to the same reference cavity in the experiment. $850$ \& $854$ Raman Ramsey fringe between $\ket{D_{3/2},m=-1/2}$ and $\ket{D_{5/2},m=-5/2}$ is measured, fitting to a Gaussian decay gives a Raman coherence time of $1019 \pm 34\,\mu$s. Due to the narrow linewidth of the $850\,\text{nm}/854\,\text{nm}$ laser, the Raman transition has a long coherence time of $\sim1\,$ms (see Fig.~S4(b)), so that the optical phases of the beams remain constant over the interval of $\sim10\,\mu$s between two successive Raman pulses. Thus the optical phases are cancelled and the relative phase can be precisely controlled by setting the phases of waveforms in the radio frequency band. The main limitation to the coherence time is the laser phase noises originating from (i) imperfect laser linewidth due to limited cavity finesse ($\sim50000$) and locking electronics (servo bump in diode lasers) and (ii) fiber noise induced by the long transmission fibers ($\sim20\,$m) from the lasers to the ion.

\begin{figure}[htbp]
    \centering
        \includegraphics[width=0.8\textwidth]{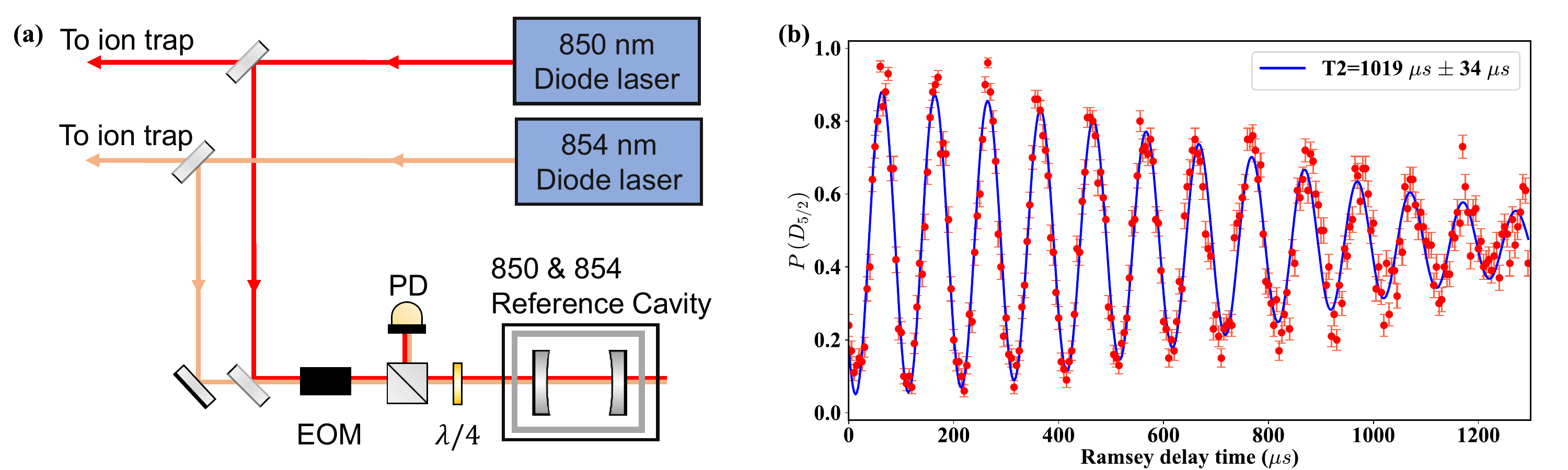}

    \caption{$850\,$nm \& $854\,$nm laser locking scheme and Raman Ramsey. (a) Locking schemes for two lasers. (b) Raman Ramsey fringe measurement.}

\end{figure}

\begin{figure}[htbp]
    \centering
    \includegraphics[width=0.6\linewidth]{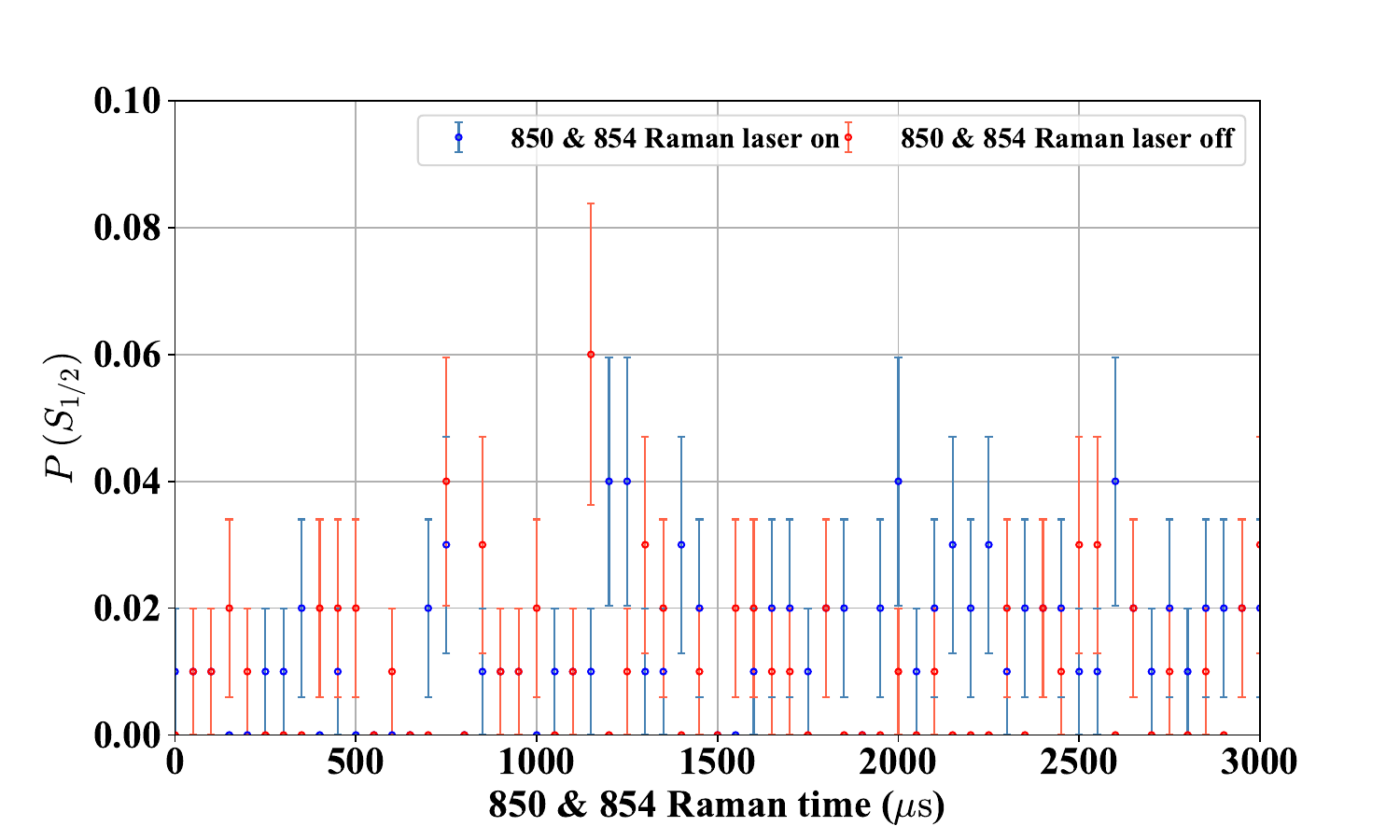}
    \caption{Influence of $850\,$nm \& $854\,$nm Raman transition to memory qubit population. We measure the unwanted scattering probability of Raman transitions with variable Raman interaction time.}

\end{figure}

In the experiment,  $850$ \& $854$ Raman transitions transfer qubits between two subspaces. The Raman pulse should not influence the memory qubit due to polarization and frequency mismatch. We prepare memory qubit to $\ket{+}=\frac{\ket{\uparrow_m}+\ket{\downarrow_m}}{\sqrt{2}}$ and measure population of $\ket{S_{1/2}}$ to detect the unwanted scattering, with $850$ \& $854$ Raman laser on or off. As shown in Fig.~S5, we cannot observe any notable difference between two cases in a duration of approximately $1000$ $\pi$-pulses. The population decay happens with a chance $<10^{-5}$ for each $\pi$-pulse.

\subsection{Section S5. Polarization preserving QFC and spectrum filtering}

Our frequency conversion device to telecom C-band is realized via difference frequency generation (DFG) process in a periodically poled lithium niobate (PPLN) waveguide. The $1.5\,$cm waveguide is placed at the center of a Sagnac type interferometer to achieve polarization preserving. Temperature of the waveguide is set to $71.5\, ^\circ \text{C}$ to meet the best quasi-phase-matching condition.

\begin{figure}[h]
	\centering
	\includegraphics[width = 0.9 \linewidth]{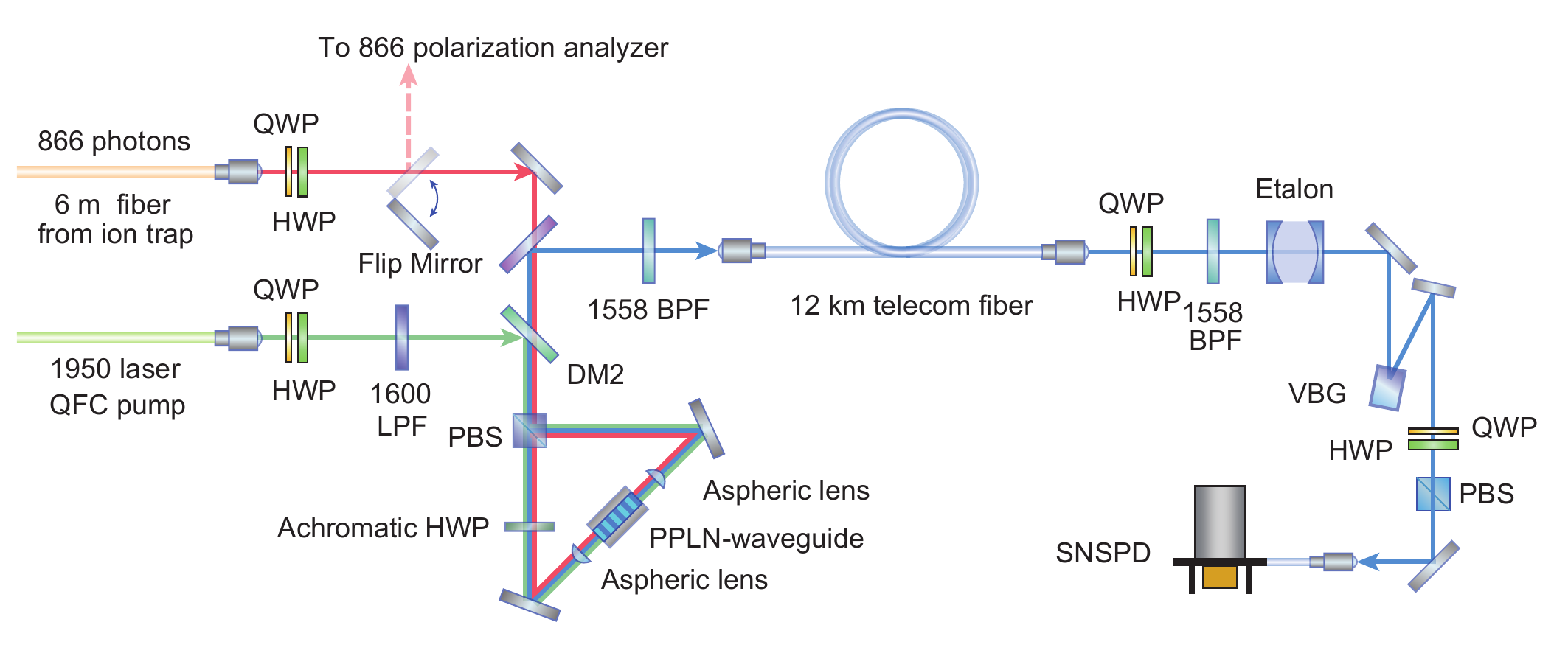}
        \caption{The PPQFC and spectrum filtering setup. The setup diagram for the quantum frequency conversion module.}

\end{figure}

As illustrated in Fig.~S6, the $866\,$nm signal photons are mixed with a strong $1950\,$nm CW pump field by a dichroic mirror and then sent into the Sagnac interferometer. Here the beam is divided into two arms in accordance with its polarization on the PBS (customized for three wavelengths). One arm with H polarizing is then flipped to V polarizing on an achromatic half wave plate to meet the phase matching polarization of the PPLN-waveguide. We place two aspheric lenses symmetrically to couple $866\,$nm photons and $1950\,$nm pump field together into PPLN-waveguide on dual ends with about $80\%$ coupling efficiency. Both the aspheric lenses and our waveguide have customized anti-reflective coatings for all three wavelengths. After the $866\,$nm signal photons are converted to $1558\,$nm in the waveguide, they travel along the rest parts of the interferometer and coherently combined to $1558\,$nm polarization qubits on the PBS. Finally converted photons are picked out by a dichroic mirror and collected into a $12\,$km G652D telecom-fiber with $80\%$ coupling efficiency. The long fibers at $1\,$km and $12\,$km are spooled fibers placed in the lab. No shielding or thermal stabilization is applied.

\begin{figure}[]
	\centering
	\includegraphics[width=0.7\linewidth]{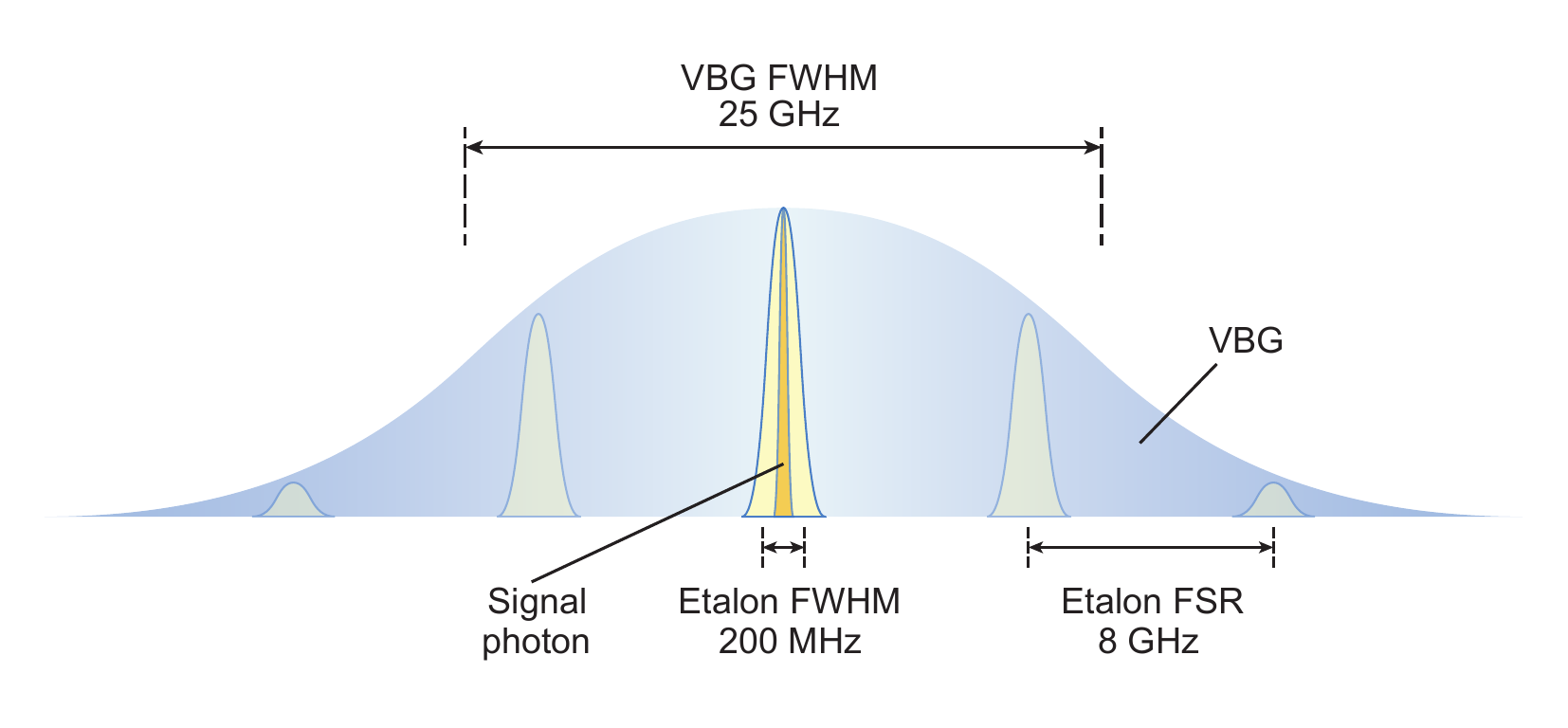}
        \caption{Narrow spectrum transmission window. The transmission spectrum for different filters and the photon linewidth.}

\end{figure}

To suppress the background noise and reach a high signal-background ratio, several types of spectrum filters are utilized, as shown in Fig.~S7. The $1950\,$nm pump laser is filtered by a long pass filter to eliminate its amplified spontaneous emission (ASE) at $1558\,$nm. In our wavelength selection, the prominent background noise in the QFC process comes from anti-Stokes Raman scattering of strong $1950\,$nm pump field. The measured Raman scattering background at $1558\,$nm is about $5\,$kHz/nm with $1.1\,$W $1950\,$nm pumping, which makes a narrow-band spectral filtering system for converted photons necessary. Here in our setup,  two band pass filters (FWHM $5\,$nm, Center $1558\,$nm) first remove the major part of the residual $1950\,$nm pump laser and other stray light from $1558\,$nm converted photons. Then the photons subsequently pass a highly stable filtering etalon (FSR $8\,$GHz, FWHM $200\,$ MHz, $90\%$ transmission) and a volume Bragg grating (VBG, FWHM $25\,$GHz, $98\%$ reflection efficiency). This combination determines a $200\,$MHz narrow spectrum transmission window which suppresses the Raman background to $8\,$Hz under $1.1\,$W $1950\,$nm CW pumping. Taking the dark count rate of SNSPD into consideration, the total background noise is $18\,$Hz and the signal background ratio is up to $22$.

\subsection{Section S6. Considerations on frequency conversion }

To determine whether an ion species is suitable for building a metropolitan-scale ($>10\,$km) quantum network, the decisive factor is whether there exists an atomic transition which can be converted to telecom band with high efficiency and low noise with current technology. As an example, here we compare the feasibility of frequency conversions of the $866\,$nm photon emitted from a $^{40}\text{Ca}^+$ ion and $370\,$nm photon emitted from a $^{171}\text{Yb}^{+}$ ion~\cite{huangyy}.

The decisive advantage of $^{40}\text{Ca}^+$ over $^{171}\text{Yb}^{+}$ is that the $866\,$nm transition in $^{40}\text{Ca}^+$ can be converted to telecom C band with high efficiency and low noise, but the $370\,\text{nm}$ photon in $^{171}\text{Yb}^{+}$ cannot. For achieving a high conversion efficiency with the current frequency conversion technology, difference frequency generation (DFG) on a periodically poled lithium niobate (PPLN) waveguide is usually employed. However, this technology requires a short poling period of the PPLN waveguide due to the strong dispersion in the lithium niobate material, to fulfill the quasi-phase-matching requirement in DFG. This dispersion is especially strong in the UV region, which results in a prohibitively short poling period. The poling period for converting a $370\,$nm photon is about $2.2\,\mu$m, which is beyond the current manufacturing precision of $2.5$-$3\,\mu$m for PPLN manufacturers. For $866\,$nm, the corresponding poling period is about $24\,\mu$m, which is easy to manufacture and can achieve a high conversion efficiency (about $40\%$ in our work, over $80\%$ has been achieved in a similar wavelength at $780\,$nm photon in~\cite{bock}).

Another consideration in the frequency conversion is that the generated noise in the DFG process should be low so that one can guarantee a high signal-to-noise ratio in the converted photon. This usually requires the pumping laser to have a longer wavelength than the converted photon, to avoid the strong noise induced by the unwanted spontaneous parametric down-conversion (SPDC) process in the PPLN waveguide. In our case, the pumping laser at $1950\,$nm cannot induce the strong SPDC noise photon near the wavelength of the converted $1558\,$nm photon, as the SPDC noise photon has a longer wavelength than the pumping laser. However, to convert the $370\,$nm photon to $1550\,$nm photon, the pumping laser is around $486\,\text{nm}$, and this will induce strong SPDC photon emission around $1550\,$nm, which would significantly reduce the signal-to-noise ratio of the converted photon. There are also some attempts to tackle the SPDC problem by using a two-stage conversion method~\cite{quraishi}. This method includes the first stage to convert the atomic transition to an intermediate wavelength, and then convert the intermediate wavelength to telecom band via second frequency conversion. Via this method, one can guarantee that the pumping laser has a longer wavelength over the converted photon in each stage, thus the SPDC noise photon can be solved. However, this method will induce lower overall conversion efficiency as the efficiencies of the two stages need to be multiplied. Besides, the difficulty of how to convert $370\,$nm photon to a longer wavelength in the first stage is still not solved, as the $370\,$nm photon cannot be efficiently converted due to the required short poling period. Therefore, currently there is no demonstrated method to convert a $370\,$nm photon to telecom C band with high efficiency and low noise. This is the decisive reason of using $^{40}\text{Ca}^+$ for building a long-distance quantum network in this work.

As explained above, the biggest and decisive problem for $370\,$nm photon from a $^{171}\text{Yb}^{+}$ ion is in the frequency conversion. If the frequency conversion from $370\,\text{nm}$ could be implemented with high efficiency and low noise in the future, we think $^{171}\text{Yb}^{+}$ would be a better choice over $^{40}\text{Ca}^+$. However, under current conversion techniques, we choose $^{40}\text{Ca}^+$ for the task of long-distance quantum network.

\renewcommand{\arraystretch}{1.5}
\renewcommand\tabularxcolumn[1]{m{#1}}
\begin{table}[h!]
  \begin{center}
  \small
    \begin{tabularx}{0.9\textwidth} {
    >{\centering\arraybackslash}X
    >{\centering\arraybackslash}X
    >{\centering\arraybackslash}X
    >{\centering\arraybackslash}X
    >{\centering\arraybackslash}X
    >{\centering\arraybackslash}X
    >{\centering\arraybackslash}X}
    \hline
    \hline
     &  $3\,$m case & $3\,$m case with future improvements &  $1\,$km case & $1\,$km case with future improvements & $12\,$km case & $12\,$km case with future improvements \\[1pt]
    \hline
    Branching ratio and weight from CG-coefficient  &  $6\times\frac{2}{3}=4\%$ &  $4\%$ & $4\%$  & $4\%$ & $4\%$ & $4\%$\\
    picosecond pulse excitation probability & $95\%$ & $95\%$ & $95\%$ & $95\%$ & $95\%$ & $95\%$\\
    Collection efficiency of objective & $6\%$  &  $9\%$ & $6\%$  &  $9\%$ & $6\%$  &  $9\%$\\
    Single mode fiber coupling efficiency & $32\%$  &  $50\%$ & $32\%$  &  $50\%$ & $32\%$  &  $50\%$\\
    Wavelength conversion efficiency (end-to-end) &  /   & /  &  /   & / & $11\%$  & $50\%$ \\
    Flying qubit transmission in fiber &  $80\%$ &   $90\%$ & $32\%$ & $36\%$  & $40\%$  & $50\%$ \\
    Other optical elements transmission (mirrors, lenses, fibers, vacuum chamber window, etc.)  &  $75\%$  & $90\%$  & $75\%$  &  $90\%$  & $31\%$ & $75\%$  \\
    Detector efficiency & $40\%$  & $90\%$   & $40\%$  & $90\%$  &  $65\%$  & $90\%$ \\
    Attempting rate & $264\,$kHz  &  $300\,$kHz & $49\,$kHz   & $80\,$kHz & $5\,$kHz &  $80\,$kHz (with multiplexing enhancement, see Section S8)  \\
    \hline
    Success rate & $46\,$Hz & $374\,$Hz & $3.4\,$Hz  & $40\,$Hz & $0.032\,$Hz & $23\,$Hz \\
    \hline
    \hline
    \end{tabularx}
  \end{center}
  \caption{Budget of ion-photon entanglement rate for current experiment and future improvement.}
\end{table}

\subsection{Section S7. Success rate and fidelity budget for ion-photon entanglement}

In this section, we decompose the total success rate of ion-photon entanglement into different parts as listed in Table S1. We also list the error budget for the fidelity.

In the future, it is possible to replace the current NA$\,=\,$$0.52$ objective to an objective with a higher NA$\,=\,$$0.6$~\cite{monroe_barium,oxford sr}, to improve the collection efficiency of the objective from $6\%$ to $9\%$. We can also improve the coupling efficiency from the light collected by the objective into the single mode fiber from the current $32\%$ to $50\%$. In addition, we can replace the uncoated single mode fiber with coated fiber to improve the transmission efficiency. We can also improve the efficiency of other optical elements by better coating and removing unnecessary fiber jumpers in current experiment. Futhermore, we can improve the attempting rate by removing unnecessary waiting time and optimizing intermediate sympathetic cooling, to $300\,$kHz in $3\,$m case, $80\,$kHz in $1\,$km case. We can also improve the attempting rate of $12\,$km fiber case to $80\,$kHz by the multiplexing enhancement described in Section S8. Currently in our experiment, the end-to-end wavelength conversion efficiency (including all the optics for spectrum filtering) is about $11\%$, which includes the efficiency of $38\%$ in the PPLN waveguide and another $30\%$ efficiency of the filtering optics. In the future, it is possible to improve the end-to-end conversion efficiency to $50\%$ as the current state of the art has an efficiency of $57\%$~\cite{bock}.

\begin{table}[h!]
  \begin{center}
  \small
    \begin{tabularx}{0.6\textwidth} {
    >{\centering\arraybackslash}X
    >{\centering\arraybackslash}X
    >{\centering\arraybackslash}X
    >{\centering\arraybackslash}X}
    \hline
    \hline
     &  $3\,$m case & $1\,$km case  &  $12\,$km case  \\[1pt]
    \hline
    Detector dark count and noise photon  &  $0.1\%$ &  $0.2\%$ & $3.4\%$  \\
    State detection & $0.5\%$  &  $0.5\%$ & $0.5\%$  \\
    $729\,$nm rotation pulse & $1.0\%$  &  $1.4\%$ & $1.5\%$  \\
    $850\,$nm / $854\,$nm Raman transition &  $0.8\%$   & $1.2\%$  &  $1.2\%$  \\
    $866\,$nm / $866\,$nm Raman transition &  $3.8\%$   & $5.5\%$  &  $5.7\%$ \\

    \hline
    Total infidelity & $6.2\%$ & $8.8\%$ & $12.3\%$   \\
    \hline
    \hline
    \end{tabularx}
  \end{center}
  \caption{Budget of infidelity in ion-photon entanglement.}
\end{table}

Here we also list all the infidelities in ion-photon entanglement in Table S2. The infidelity induced by detector dark count and noise photon is low in $3\,$m and $1\,$km case. In $12\,$km case, the signal-to-noise ratio of $22$ induced by the imperfect spectral filtering in wavelength conversion results in an infidelity of $3.4\%$. Here we assume the ion-photon state heralded by the noise photon$/$dark count is a fully mixed state. This infidelity can be resolved by better spectrum filtering in the wavelength conversion in future. The infidelity induced by $729\,$nm rotation pulse is $1.0\%$ in $3\,$m case in which the protocol time is $40\,$ms. For the $1\,$km and $12\,$km case, the protocol time is $200\,$ms. The longer sequence time influences the quality of the active stabilization of laser intensity, which causes higher infidelities in the $729\,$nm, $850\,$nm$/$$854\,$nm Raman and $866\,$nm$/$$866\,$nm Raman pulses for $1\,$km and $12\,$km cases. This can be improved in the future by more frequent stabilization during the sequence, but currently it influences the fidelity of ion-photon entanglement in our experiment. Due to the current low photon generation rate in $12\,$km case, the accuracy of the phase calibration in $729\,$nm rotation and $866\,$nm$/$$866\,$nm Raman pulse (see Section S9) are also influenced, which further increases the infidelities in these two operations in the $12\,$km case. With improved photon generation rate in the future, this effect can be suppressed.

\subsection{Section S8. Heralded ion-ion entanglement }

\begin{figure}
	\centering
	\includegraphics[width=0.9\linewidth]{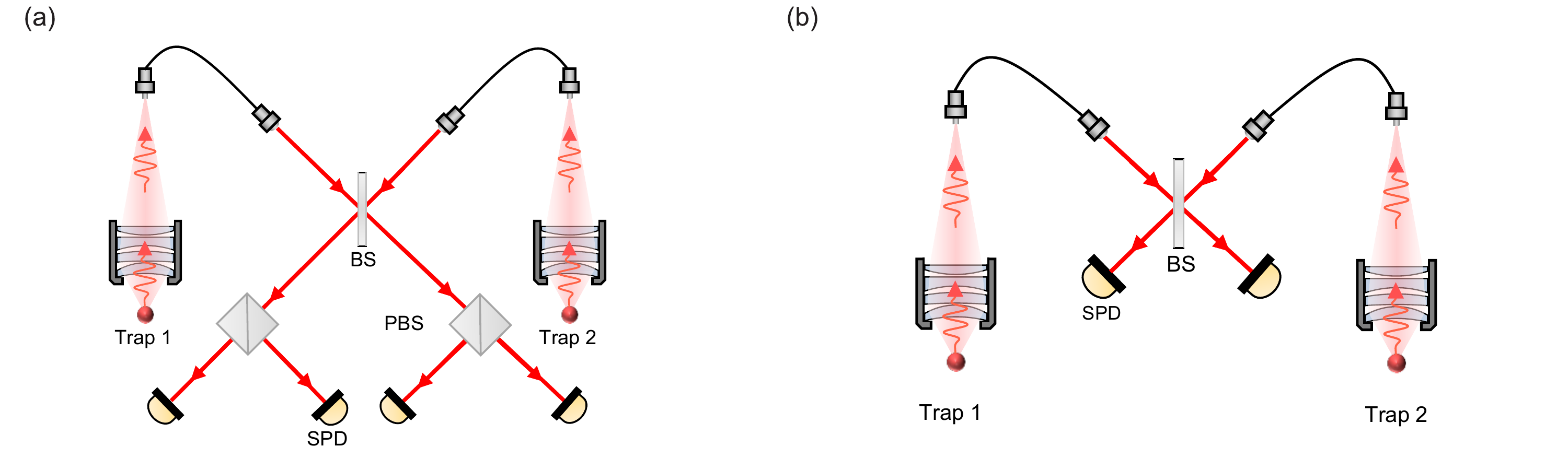}
        \caption{Two heralding schemes for ion-ion entanglement. (a) Ion-ion entanglement can be heralded by two-photon Bell state measurement (BSM), by recording a two-photon coincidence event. (b) Ion-ion entanglement can be heralded by single-photon interference, by recording a single photon on either detector after the beamsplitter (BS).}
\end{figure}

In this section we discuss the schemes to generate heralded ion-ion entanglement in the future with two such setups demonstrated in this work. We also discuss several methods to improve the generation rate of heralded ion-ion entanglement.

First, we can use the two-photon Bell state measurement (BSM) to generate ion-ion entanglement (see Fig.~S8(a)), similar to the recent experiments as demonstrated in~\cite{oxford sr,monroe_barium}. The coherence time of $323\,$ms is far longer than the required waiting time of $2L/c=120\,\mu$s in each entangling attempt ($L=12\,$km is the fiber length), and the high SNR of $20$ (can be further improved in the future, see paragraphs below) can also guarantee a high fidelity of the generated ion-ion entangled state.

For the lab-scale case ($3\,$m), we can also switch to collecting $397\,$nm photon instead of the $866\,$nm photon, with only very slight change in the experimental setup, while keeping all other things the same as in this experiment. Collecting $397\,$nm photon is advantageous over $866\,$nm photon in lab-scale quantum network, as the $866\,$nm transition has a low branching ratio $\sim6\%$. Note that $397\,$nm transition is also in the communication qubit subspace as $866\,$nm photon, thus we can keep all the operations and protocol the same after switching to $397\,$nm photon. The only thing need to change is the objective (even the objective does not need to change if it is designed for $397\,$nm and $866\,$nm at the same time) and the single mode fiber. Thus we can switch to $397\,\text{nm}$ with very small cost. By mediating the ion-ion entanglement via two-photon BSM with $397\,$nm photon, together with the improvement of the attempting rate from $264\,$kHz to $1\,$MHz, the heralded entanglement generation rate can be improved to the same scale of the state of the art $\sim100\,$Hz~\cite{oxford sr,monroe_barium}, which is faster than the memory decoherence rate in this experiment. However, it is worthy to note that this scheme can only be used in a lab-scale quantum network as the UV photon is used.

For the $1\,$km and $12\,$km cases, we can employ the single photon interference scheme~\cite{type I} as shown in Fig.~S8(b). Single photon scheme can achieve a much faster entangling rate than the two-photon BSM. Single photon interference has been demonstrated in many experiments before~\cite{bao3nodes,2021nature,hanson_distillation,slodicia,hanson long}, either in a lab-scale quantum network or metropolitan-scale quantum network. The main difficulty in single photon interference is the stabilization of the relative phase of the optical paths from each node to the beamsplitter in the middle detection station, which has been experimentally realized in different systems, especially in trapped-ion system~\cite{slodicia} and the NV center system~\cite{hanson_distillation,hanson long} which is similar with the ionic system in this experiment. With single photon interference, the ion-ion entangling rate will be close to the ion-photon entangling rate. Thus by using single photon scheme, we can achieve an ion-ion generation time of roughly $21\,$ms ($46\,$Hz) in a fiber length of $3\,$m, and ion-ion entanglement generation time of $294\,$ms ($3.4\,$Hz) over a fiber of $1\,$km. In the future, the entangling time can be further improved to $2.7\,$ms ($374\,$Hz) in $3\,$m case and $25\,$ms ($40\,$Hz) in $1\,$km case, as shown in Table S1. With the ion-ion entanglement generation time shorter than the memory coherence time of $323\,$ms, we can implement multi-node quantum network based on many setups demonstrated in this work. Note that achieving a memory coherence time longer than heralded atom-atom entanglement generation time is an ultimate objective, which has only been achieved in a lab-scale quantum network~\cite{hansondelivery,monroe}. This requirement has not been achieved over a fiber of 1km in any physical platforms so far.

Other schemes to accelerate the remote ion-ion entanglement generation are multiplexing enhancement and cavity enhancement. We estimate the time cost for each entangling attempt can be improved to about $10\,\mu$s via multiplexing enhancement~\cite{enhancement, lanyon multiplexing, haffner multiplexing} in the future, which is $12$ times faster than the $120\,\mu$s in this experiment. We expect an average attempting rate of $80\,$kHz can be achieved in the future. We can also improve the overall efficiency in the optical channel by roughly $45$ times, as shown in Table S1. With these improvements, we can already achieve a remote ion-photon entangling time (equal to the ion-ion entangling time via single photon interference, see Section S15) of $43\,$ms over $12\,$km fiber in the future, which is shorter than the memory coherence time $323\,$ms. The signal-to-noise ratio (SNR) of the flying qubit can also be improved by the enhancement of efficiency. Note that the remote ion-ion entanglement generation rate is also at the same level if single photon interference is used. The entanglement generation rate can be further improved if cavity enhancement is also employed. Note that a high-finesse cavity designed at $866\,$nm~\cite{866 cavity} will be perfectly compatible with the dual-type encoding in this work.

Another consideration is that, in a future functional large-scale quantum network, it is advantageous to have a high remote entanglement distribution rate between different nodes, for the applications such as distributed quantum computing or quantum repeater. If the generation rate of the remote entanglement is too slow, the remote gate speed cannot match the clock speed of the local quantum computer which is about $10\,$kHz (two-qubit gate time is about $100\,\mu$s in trapped-ion quantum computer) in a distributed quantum computing task. For a quantum repeater, a faster entanglement distribution time is also favored, to beat the direct photon transmission in a quantum key distribution task ($\sim40\,$Hz over $500\,$km, see~\cite{zhou lai}). Too slow entanglement distribution rate will significantly compromise the functionality of a quantum network. Therefore, although it is both important to improve the remote entangling rate and the memory coherence time, we think it is more urgent to improve the remote entangling rate to demonstrate a functional quantum network which can exceed the current ability in quantum information processing tasks. With a much improved remote entangling rate in the future, the $323(19)\,$ms coherence time is not short. For example, a typical quantum computing task in trapped-ion quantum computer won't take so long. For a quantum repeater, $323\,$ms coherence time is much longer than the required entanglement generation time in each elementary link ($43\,$ms for $12\,$km case, see Table S1 and the discussions above) if the entanglement generation rate can be greatly improved in the future. Thus the $323\,$ms coherence time can already support a metropolitan quantum repeater, if the entanglement generation rate can be significantly improved in the future.

Here we also investigate the reduction in success rate due to memory qubit decay in ion-ion entanglement. Here we define the criteria of success, which include: (1) successful generation of ion-ion entanglement during the $40\,\text{ms}/200\,\text{ms}/200\,\text{ms}$ attempting periods in the $3\,\text{m}/1\,\text{km}/12\,\text{km}$ cases., and (2) the spontaneous decay does not happen in memory qubits in both of the two nodes. Then the success rate reduction due to the memory qubit decay is $9.6\%/39.7\%/39.7\%$ for the $3\,\text{m}/1\,\text{km}/12\,\text{km}$ cases.

With the above improvements, we expect a remote entanglement generation time of $43\,$ms over a metropolitan fiber can be achieved. As this coherence time is much shorter than the memory coherence time $323\,$ms, we think many advanced applications such as three-node (or even more nodes) metropolitan-scale quantum repeater~\cite{duan rmp} based on entanglement connection of asynchronously entangled elementary links, distributed quantum computing~\cite{duan pra} of metropolitan scale, entanglement distillation~\cite{lukin_siv,hanson_distillation} in a quantum network or other prominent applications can be achieved.

\subsection{Section S9. Implementation of $866\,\text{nm}/866\,\text{nm}$ Raman }

\begin{figure}
	\centering
	\includegraphics[width=0.9\linewidth]{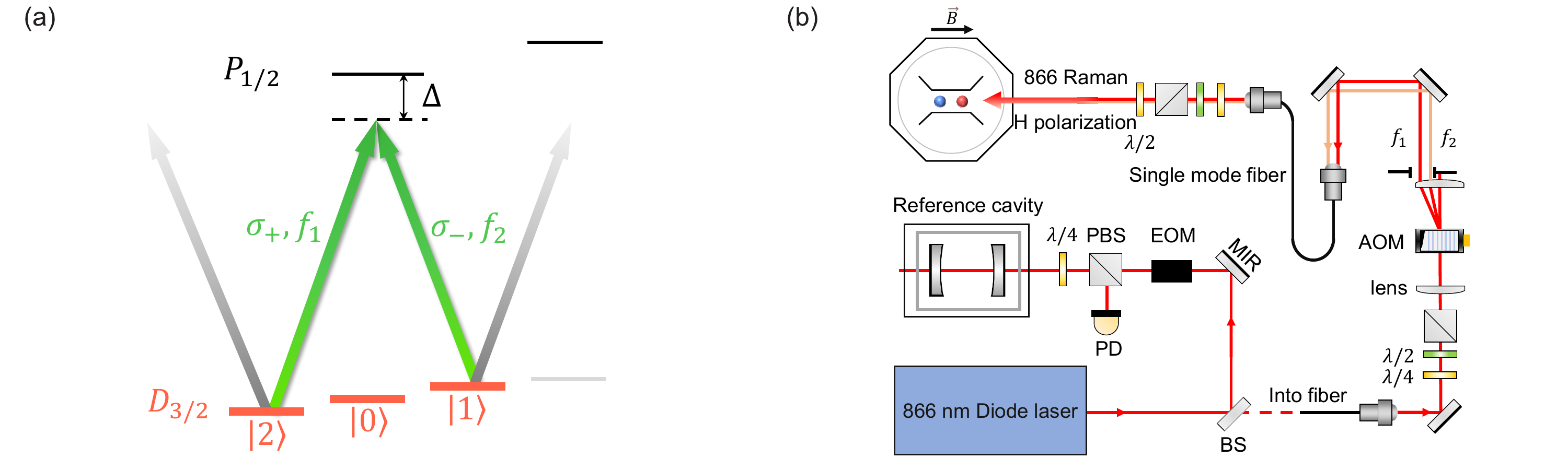}
        \caption{The implementation of $866\,\text{nm}/866\,\text{nm}$ Raman. (a) Only the two components ($\sigma^+$, $f_1$) and ($\sigma^-$, $f_2$) out of the four combinations of frequency and polarization can bridge a Raman transition between $\ket{1}$ and $\ket{2}$. (b) The optical setups of the $866\,\text{nm}/866\,\text{nm}$ Raman. The two frequency components are two different $1$st-order diffractions from the same AOM and laser beam.}
\end{figure}

\begin{figure}
	\centering
	\includegraphics[width=\linewidth]{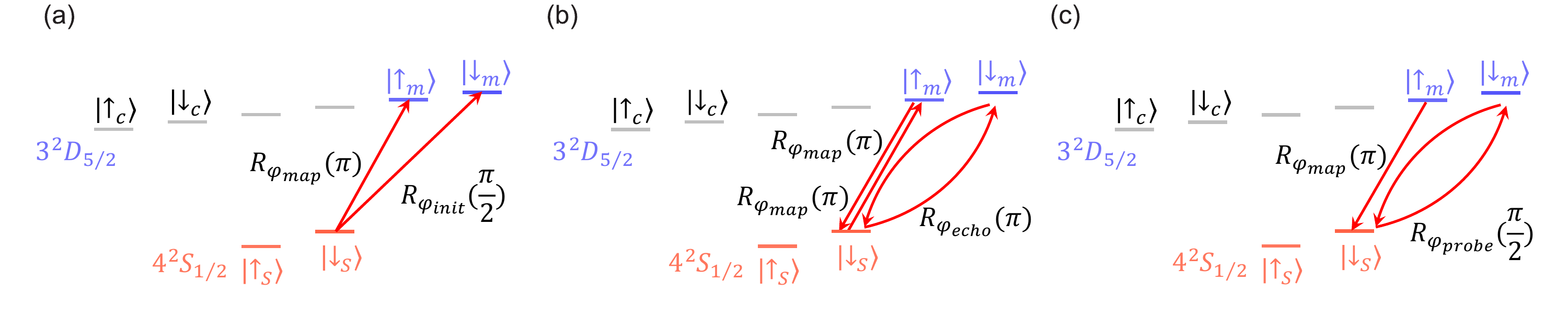}
        \caption{The operations on the memory qubit. (a) We prepare the stored qubit $\frac{\ket{\uparrow_m}+\ket{\downarrow_m}}{\sqrt{2}}$ by a $729\,$nm $\frac{1}{2}\pi$-pulse and a $\pi$-pulse. (b) Spin echo via three $729\,$nm $\pi$-pulses. (c) State verification by a $\pi$-pulse and a $729\,$nm $\frac{1}{2}\pi$-pulse.}
\end{figure}

The rotation axis in $866\,\text{nm}/866\,\text{nm}$ Raman pulse is fixed by two requirements. The first requirement is that we keep the time interval between the photon click event and the beginning of the Raman pulse a constant value. By fixing this evolution time, we can guarantee the phase accumulated in the Larmor precession to be constant in each experiment. The second requirement is that we generate the two $866\,$nm Raman beams (with different circular polarizations and different frequencies) from the same laser beam, by adding two frequency tones on the same AOM (see Fig.~S9(b)). The two first order diffractions are collected into the same single-mode fiber and further sent to the ion. In addition, the polarizations of the two first order diffractions are set to H polarization at the ion, thus both frequency components have two circular polarizations. However, due to frequency and polarization selection, only two components can bridge the Raman transition between $\ket{1}$ and $\ket{2}$, out of the total four components (see Fig.~S9(a)). In this way, the optical phase of the $866\,$nm laser is canceled as the two $866\,$nm Raman beams are two components from the same laser. The rotation axis of the Raman transition is thus only determined by the RF signals driving the AOM. Here we use the photon click events to trigger the two-tone RF signals stored in an AWG, to guarantee the accumulated phase and the rotation axis are stable. For the detailed calibration, we do the following two steps to determine $866\,\text{nm}/866\,\text{nm}$ Raman phase and pulse duration.

\begin{enumerate}
    \item We set the duration of $866\,\text{nm}/866\,\text{nm}$ Raman transition roughly to be a $\frac{2}{3}\pi$  pulse and scan the relative phase between the two RF frequency components in the AWG output and measure $\ket{1}$ population vs this phase to get the phase for the correct rotation axis.
    \item With the phase fixed to the result in the first step, we then scan the pulse duration of $866\,\text{nm}/866\,\text{nm}$ Raman transition and measure $\ket{1}$ population to determine the correct pulse duration, as shown in Fig.~2(d) in the main text.
\end{enumerate}

We update the pulse duration in step $1$ according to the result of step $2$ and iterate the above process several times to get the correct phase (rotation axis) and pulse duration of the $866\,\text{nm}/866\,\text{nm}$ Raman transition.

\subsection{Section S10. Implementation of spin echo}

In this section we demonstrate how to implement the state preparation, spin echo, and state verification on memory qubit in the sequence demonstrated in Fig.~3(a). Here we assume the stored quantum information is $\frac{\ket{\uparrow_m}+\ket{\downarrow_m}}{\sqrt{2}}$ as an example. As illustrated in Fig.~S10(a), we initialize the state of the memory qubit by a $729\,$nm $\frac{\pi}{2}$-pulse followed by a $729\,$nm $\pi$-pulse. The relative phase between them is controlled by the relative phase between the two $729\,$nm pulses. Then after a storage time of $\tau$, we apply the spin echo by three steps: (i) swapping $\ket{\uparrow_m}$ and $\ket{\downarrow_s}$ by a $729\,$nm $\pi$-pulse, (ii) swapping $\ket{\downarrow_s}$ and $\ket{\downarrow_m}$ by a $729\,$nm $\pi$-pulse, and (iii) swapping $\ket{\uparrow_m}$ and $\ket{\downarrow_s}$ by a $729\,$nm $\pi$-pulse, as shown in Fig.~10(b). After these three pulses, the basis $\ket{\uparrow_m}$ and $\ket{\downarrow_m}$ are swapped, and the spin echo is completed. Finally, to verify the stored state after storage time of $2\tau$, we apply a $729\,$nm $\pi$-pulse followed by a $729\,$nm $\frac{\pi}{2}$-pulse, as shown in Fig.~10(c). The state verification is finalized by a fluorescence detection. Note that in all these operations, the $729\,$nm pulses are applied via an addressed $729\,$nm beam, thus these operations will not influence the communication qubit.

We do not arrange photon generation attempts during or closely before the spin echo ($120\,\mu$s before or after the spin echo). Thus the spin echo operation will not influence the communication qubit.

\subsection{Section S11. Implementation of entanglement swapping}

\begin{figure}
	\centering
	\includegraphics[width=\linewidth]{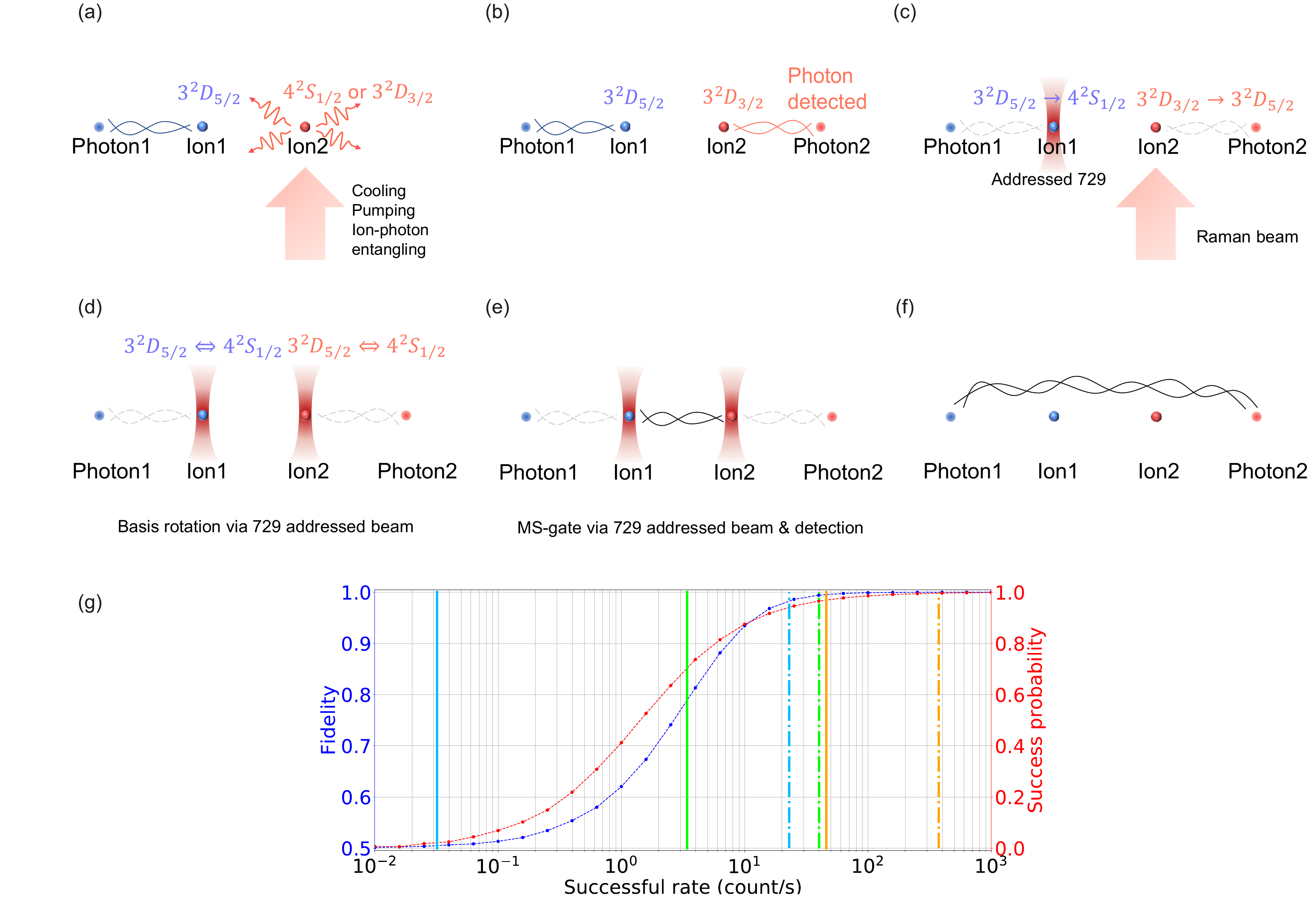}
        \caption{The implementation of entanglement swapping. (a, b) Generate heralded ion-photon entanglement on ion 2. (c, d) State shelving and preparation for MS gate. (e, f) Entanglement between photon 1 and photon 2 is established after MS gate and state detection. (g) The success probability (red curve) and photon-photon entanglement fidelity after swapping (blue curve) as a function of the ion-photon entanglement generation rate, under the condition that the coherence time is $323\,$ms and the lifetime is $790\,$ms. The solid orange$/$green$/$blue vertical line represents the ion-photon entangling rate of $3\,$m$/$$1\,$km$/$$12\,$km case, respectively. The dashed orange$/$green$/$blue vertical line represents the improved ion-photon entangling rate (see Table S1) of $3\,$m$/$$1\,$km$/$$12\,$km case in the future, respectively. }
\end{figure}

In this section we demonstrate how to implement the entanglement swapping in our two-ion node, by the operations demonstrated in this work and an additional $729\,$nm MS (Molmer-Sorensen) gate. The core tasks of a quantum network include: (i) the generation of remote entanglement between distant nodes in a crosstalk-free way, and (ii) the local entanglement swapping between two ions in a single node to extend the entanglement to a longer distance. For task (i), this has been demonstrated in this work. To implement task (ii), we can employ the coherent conversions between communication and memory subspaces demonstrated in this work and standard gates based on addressed $729\,$nm beams. Note that addressed $729\,$nm beam controlled by crossed AODs is already demonstrated in this experiment. Each ion can be converted between memory qubit encoding and communication qubit encoding by the coherent conversion operations when we perform the quantum logic operations. This is because the spontaneous emission crosstalk is negligible during quantum logic operations, thus we don't need to stick to the spectrally separated encoding during the quantum logic operations. High fidelity gate itself can guarantee the spontaneous emission is negligible, as the gate will fail if spontaneous emission happens. (For state detection, we still need to shelve other ions to metastable level.) The crosstalk in the quantum logic operations can be suppressed by improving the addressing system, which is same as in any trapped-ion quantum computer.

\begin{enumerate}
    \item We start from the point that ion 1 has been entangled with photon 1 for simplicity, as shown in Fig.~S11(a). Note that photon 1 can also be another ion in different network node, if ion-ion entanglement is heralded. In this stage, ion 1 is encoded in the memory qubit subspace $D_{5/2}$. Ion 2 is in the communication qubit subspace and repetitive cooling, pumping, and ion-photon entangling operations are applied to ion 2, to herald another ion-photon entanglement.
    \item If a photon is detected in the ion-photon entangling of ion 2, ion 2 and photon 2 are entangled, as shown in Fig.~S11(b). The cooling, pumping, ion-photon entangling operations halts after the entanglement is heralded.
    \item Then, we map the qubit stored in ion 1 from $D_{5/2}$ level to $S_{1/2}$ level via an addressed $729\,$nm beam, as shown in Fig.~S11(c). After that, we use the $850\,\text{nm}/854\,\text{nm}$ and $866\,\text{nm}/866\,\text{nm}$ Raman pulses to convert the state of ion 2 from $D_{3/2}$ level to $D_{5/2}$. Because ion 1 is in $S_{1/2}$, the global $866\,\text{nm}/866\,\text{nm}$ and $850\,\text{nm}/854\,\text{nm}$ Raman will not influence ion 1.
    \item We further map the state of ion 1 and ion 2 to the basis ($\ket{\downarrow_s}$ and $\ket{\downarrow_m}$ for example) for the following MS gate, by two successive addressed $729\,$nm pulses, as shown in Fig.~11(d).
    \item Then we perform the MS gate via $729\,$nm addressed beams followed by a fluorescence detection on both ions, as shown in Fig.~S11(e).
    \item After the entanglement swapping, photon 1 and photon 2 are entangled, as shown in Fig.~S11(f).
\end{enumerate}

One can see that, with the operations demonstrated in this work and an MS gate widely demonstrated before~\cite{lanyon_connection, MS1, MS2, MS3}, we can achieve the entanglement generation and swapping, which are the core tasks in a quantum network node. More advanced operations such as error correction can also be performed in this way.

In addition, we also investigate the influence of the limited memory qubit lifetime $T_{1}$ and coherence time $T_{2}$ to the task of the memory-enhanced connection of two asynchronously generated atom-photon entanglement discussed above. Note that this application is also very similar with the entanglement connection in a three-node quantum repeater, if one replaces the generation of heralded ion-photon entanglement to heralded ion-ion entanglement on each side. There are already studies on this issue, which are investigated in detail in~\cite{hansondelivery, duan pra}. Here we also study how the fidelity and success probability vary with different ion-photon entangling rates, under the situation of our current experiment with $T_{1}=0.79\,$s and $T_2=0.323\,$s.

Here we simulate the fidelity of the remote photon-photon entanglement and the success probability versus the ion-photon entanglement generation rate. The protocol is the same as the protocol discussed above, including: (1) we first generate heralded entanglement between ion 1 and photon 1; (2) convert ion 1 to memory qubit; (3) generate ion-photon entanglement between ion 2 and photon 2; and (4) perform entanglement swapping on ion 1 and ion 2 to generate entanglement between photon 1 and photon 2 (state detection is applied before swapping to discard the events when spontaneous emission happens). Both of the spontaneous decay and decoherence in memory qubit (ion 1) can happen in step (3), which are determined by the remote entanglement generation rate, $T_1$ and $T_2$. In this protocol, the spontaneous decay influences the success probability of the protocol and the decoherence influences the fidelity of the final photon-photon entanglement. To emphasize the effect of the spontaneous decay and decoherence, here we assume an ideal case: the newly generated ion-photon entanglement has a fidelity of $1$, and all the gates (including rotation and entanglement swapping) have perfect fidelities. In this simulation, we assume the diagonal part of the density matrix of the ion-photon entangled state does not change with storage time (after discarding the spontaneous decay events by state detection), and the off-diagonal part of the density matrix suffers from a Gaussian decay with $1/e$ decay time of $T_2=323\,$ms. Here the result of the simulation is shown in Fig.~S11(g). Here we also show $6$ different cases as described in Section S7 and Table S1. One can see that, with the improvement in the ion-photon entanglement generation rate in the future, both of the fidelity and success probability can be high with $T_{1}=0.79\,$s and $T_2=0.323\,$s in our experiment.

For the experiment implemented in this paper following the sequence shown in Fig.~3a. (and results demonstrated in Fig.~3 and Fig.~4), the crosstalk of $729\,$nm laser and Raman beams are negligible. Specifically, the three Raman pulses for bases conversion after ion-photon entanglement will not influence the qubit encoded in memory basis. As described in Section S4, we estimate the crosstalk is below $10^{-5}$ per Raman pulse. The $729\,$nm laser in the qubit rotation operation also won't influence the memory qubit. We estimate the crosstalk is at the level of $10^{-6}$ with both spatial ($10^{-2}$) and frequency isolation ($10^{-4}$). Thus we can map the communication qubit to memory manifold without introducing crosstalk.

For the proposed entanglement swapping experiment in Section S11, there exists cross-talk in preparing the bases with addressed $729\,$nm beams for the following MS gate after the remote entanglement is established. However, this crosstalk error commonly exists in a trapped-ion quantum computer, and there are lots of schemes to solve this crosstalk. For the individually addressed $729\,$nm beam used in our experiment, the leaked $729\,$nm laser intensity on the neighboring ion is below $10^{-4}$, which corresponds to less than $1\%$ crosstalk to the neighboring ion when driving a resonant $\pi$-pulse rotation directly. Furthermore, there are many solutions to suppress the crosstalk. One can resolve this crosstalk by improving the performance of individual addressing~\cite{AQT}, exploiting composite gate schemes~\cite{AQT}, or shuttling the ion to a different region of the trap~\cite{quantiuum}. With these solutions, one can achieve a crosstalk less than $10^{-4}$.

\subsection{Section S12. Future improvement on attempting rate}

In this section we discuss the limiting factors for the current attempting rate of $264\,$kHz, $49\,$kHz and $5\,$kHz for the cases of $3\,$m, $1\,$km and $12\,$km fiber in our experiment, as well as the future improvement.

For the case of $3\,$m fiber, the attempting rate $264\,$kHz is determined by the AOM latency ($\sim200\,$ns), overhead such as cooling, and the redundant waiting time between different operations (for example, the unnecessary waiting time between pumping and ion-photon generation) in each attempt, as we haven't optimized the attempting rate by removing all the unnecessary waiting time in this experiment. If all the unnecessary waiting time is removed and the overhead is minimized, we expect we can reach an attempting rate at $\sim1\,$MHz~\cite{monroe_barium,oxford sr} with two photon scheme and $300\,$kHz for single photon scheme, mainly limited by the AOM latency and intermediate sympathetic cooling.

For the case of $1\,$km fiber, the attempting rate is mainly limited by the waiting time $2L/c$ in each heralded entanglement attempt, which includes the time $L/c$ for transmitting the photon over the $L=1\,$km fiber to the detector, and another $L/c$ for the heralding signal to come back through another $L=1\,$km fiber. Here $2L/c=10\,\mu$s for a $1\,$km fiber, which corresponds to a maximum attempting rate of $100\,$kHz. Here we are using an attempting rate of $49\,$kHz as we haven't optimized the attempting rate. The unnecessary waiting time and some other overhead like cooling result in an attempting rate lower than the theoretical upper limit $100\,$kHz. In the future if everything is optimized, we expect an attempting rate of $80$$\,$kHz can be achieved.

For the case of $12\,$km fiber, the $2L/c = 120\,\mu$s waiting time (corresponding to $8.3\,$kHz) is the main limit for the attempting rate. With optimized overhead and sequence, we estimate an attempting rate of roughly $8\,$kHz can be achieved. In addition, with future multiplexing enhancement, we expect an attempting rate as fast as $80\,$kHz can be reached.

\subsection{Section S13. Estimation of the influence induced by crosstalk operations}

Here we estimate the excess decoherence and population decay in memory qubit induced by the cooling, optical pumping and entanglement excitation operations on the communication qubit.

The noisy operations for generating ion-photon entanglement in our experiment include: (1) a strong $397\,$nm $10$-ps pulse which drives a $\pi$-pulse to excite the ion-photon entanglement. This beam has a Gaussian radius of about $10\,\mu$m. (2) a $290\,$ns optical pumping containing $397\,$nm  and $866\,$nm lasers to prepare the ion to the initial state $|S_{1/2},m=1/2\rangle$. (3) a $200\mu$s sympathetic EIT cooling (containing a strong $\sigma^{+}$-polarization $397\,$nm beam, a weak $\pi$-polarization $397\,$nm beam and a $866\,$nm beam) which is applied every $100$ entangling attempts. The lasers used in (2) optical pumping and (3) cooling has a large beam size which cover both ions with very similar intensities. Note that although in principle the size of these lasers can be reduced so that negligible leakage light will shine on the memory qubit, currently in our experiment these leakage light are still the dominating source of the crosstalk.

We first investigate the potential population decay in memory qubit (decay from $|D_{5/2}\rangle$  to $|S_{1/2}\rangle$ via $|P_{3/2}\rangle$) induced by the operations applied to communication qubit. We estimate the scattering rate for the memory qubit encoded in $|D_{5/2}\rangle$ manifolds via the off-resonant excitation~\cite{Foot}. The probability of exciting the ion from $|D_{5/2}\rangle$ level to $|P_{3/2}\rangle$ level is roughly
\begin{equation}
\rho = \frac{\Omega^2}{2\Omega^2+4\Delta^2}\approx\frac{\Omega^2}{4\Delta^2}
\end{equation}
where $\Omega$ is the Rabi frequency of the laser, $\Delta$ is the frequency difference between the driving laser and the $854\,$nm dipole transition from $|D_{5/2}\rangle$ level to $|P_{3/2}\rangle$ (which is much stronger than the $729\,$nm quadruple transition, thus we ignore $729\,$nm transition in the analysis). Then the spontaneous decay error in each ion-photon entangling attempt is
\begin{equation}
\epsilon = \rho\Gamma\tau=\frac{\Omega^2}{4\Delta^2}\Gamma\tau
\end{equation}
where $\Gamma$ is the spontaneous decay rate of $|P_{3/2}\rangle$ level and $\tau$ is the time duration of laser. For operation (1) $397\,$nm picosecond pulse ($\Omega=\frac{1}{2}\times\frac{1}{10\,\text{ps}}=50\,$Ghz), the spontaneous decay error is estimated to $\epsilon=5\times10^{-12}$ per excitation pulse. This corresponds to an excess decay rate of $2\times10^{-8}/$s for the attempting rate of $5\,$kHz in the $12\,$km case.

For operation (2) optical pumping, the spontaneous decay error due to $397\,$nm laser ($\Omega=10\,$Mhz) is estimated to be $6\times10^{-15}$ each operation, which corresponds to a decay rate of $3\times10^{-11}/$s. On the other hand, the spontaneous decay induced by $866\,$nm laser ($\Omega=5\,$Mhz) is $9\times10^{-12}$ per optical pumping, which corresponds to a decay rate of $5\times10^{-8}/$s.

For operation (3) intermediate cooling, the spontaneous decay due to the strong $\sigma^{+}$ $397\,$nm laser ($\Omega=10\,$Mhz) is estimated to be $4\times10^{-12}$ per operation, and corresponds to a decay rate of $2\times10^{-10}/$s (note that cooling is applied every $100$ attempts). For the $866\,$nm laser ($\Omega=1\,$Mhz), the induced decay is $3\times10^{-10}$ per operation, which corresponds to a decay rate of $1\times10^{-8}/$s.

Thus one can see that, the combined decay rate due to all the three operations is about $8\times10^{-8}/$s (see Table S3) which is negligible compared to the natural decay rate $\sim1/$s and the measured $0.79\,$s lifetime.

Here we also investigate the memory qubit decoherence induced by the operations applied to communication qubit. The first origin of the decoherence is from the population decay of $|D_{5/2}\rangle$ state which is negligible as discussed above. In the following we discuss the potential decoherence from the differential AC Stark shift experienced by the memory qubit. The effect is due to the difference in the AC Stark shifts experienced by the two states of memory qubit $|D_{5/2},m=+3/2\rangle$ and $|D_{5/2},m=+5/2\rangle$. The phase difference due to AC Stark shift in each attempt can be expressed as:
\begin{equation}
\delta\phi_{\text{AC}} = \frac{\Omega^2}{4\Delta}\tau\mathcal{P}\alpha
\end{equation}
where $\mathcal{P}$ is the coefficient relating to the laser polarization. It is noteworthy that only the $\sigma^{-}$ and $\pi$ polarization components can induce AC Stark shift as $\sigma^{+}$ polarization does not couple the memory qubit to any energy levels in the $|P_{3/2}\rangle$ manifolds. If the $\sigma^{-}$ and $\pi$ polarization components have equal strength in the driving laser, the AC Stark shift experienced by the two basis states $|D_{5/2},m=+3/2\rangle$ and $|D_{5/2},m=+5/2\rangle$ would be identical (see the CG-coefficients in $854\,$nm transition~\cite{stute thesis}). Then the differential AC Stark shift would be zero. Besides, unlike other systems where the memory qubit is encoded on two states which are several GHz apart~\cite{oxford_sr_and_ca}, the frequency difference between two Zeeman levels in our scheme is only about $\mu_{\text{B}}g\Delta \text{m} B/h=7\,$MHz. Thus in our case the differential AC Stark shift induced by the basis energy difference is small compared to the polarizations. Here another coefficient $\alpha$ represents the suppression factor of spin echo. When spin echo is applied, the two basis states are flipped in the middle thus most of the accumulated phases are cancelled. We estimate the suppression coefficient by spin echo is $\alpha=0.01$ in our experiment.

For operation (1) $397\,$nm picosecond excitation pulse, the polarization is pure $\sigma^{-}$ (as shown in Fig.~1(b)). Thus the polarization coefficient is $\mathcal{P}=40\%$. As the $1/e^2$ Gaussian radius of the laser beam is $10\,\mu$m and the separation between two ions is $9\,\mu$m, $\Omega$ is roughly reduced to $44\%$ on the memory qubit. Thus the AC Stark shift induced phase is $\frac{\Omega^2}{4\Delta}\tau\mathcal{P}=8\times10^{-6}\,$Rad per excitation without the compensation by spin echo, and $\frac{\Omega^2}{4\Delta}\tau\mathcal{P}\alpha=8\times10^{-8}\,$Rad with spin echo. The phase difference induced by the differential AC Stark shift is estimated to  $\delta\phi_{\text{AC}}\times r=4\times10^{-4}\,$Rad$/$s, where $r=5\,$kHz is the attempting rate.

For operation (2) optical pumping, the polarization coefficient of $397\,$nm pulse is $\mathcal{P}=0.01$ as this beam is in $\sigma^{+}$ polarization with negligible components in $\sigma^{-}$ and $\pi$. Thus the the phase difference induced by the differential AC Stark shift is estimated to $6\times10^{-8}\,$Rad/s. For $866\,$nm beam, $\mathcal{P}$ is estimated to be $0.05$ as it has roughly equal distribution over $\sigma^{-}$ and $\pi$. The phase difference induced by the differential AC Stark shift from $866\,$nm laser is estimated to be $6\times10^{-6}\,$Rad$/$s.

\begin{table}[h!]
  \begin{center}
  \small
    \begin{tabularx}{0.6\textwidth} {
    >{\centering\arraybackslash}X
    >{\centering\arraybackslash}X
    >{\centering\arraybackslash}X
    >{\centering\arraybackslash}X}
    \hline
    \hline
       & population decay  &  decoherence  \\[1pt]
    \hline
    $397\,$nm picosecond pulse  &  $2\times10^{-8}/$s &  $4\times10^{-4}\,$Rad/s  \\
    optical pumping  &  $5\times10^{-8}/$s & $6\times10^{-6}\,$Rad/s  \\
    sympathetic cooling & $1\times10^{-8}/$s  &  $2\times10^{-6}\,$Rad/s   \\

    \hline
    Total influence & $8\times10^{-8}/$s & $4\times10^{-4}\,$Rad/s   \\
    \hline
    \hline
    \end{tabularx}
  \end{center}
  \caption{List of population decay and decoherence on memory qubit induced by operations on communication qubit.}
\end{table}

For operation (3) intermediate sympathetic cooling, the polarization coefficient of $397\,$nm pulse is $\mathcal{P}=0.01$ as this beam is in $\sigma^{+}$ polarization with negligible components in $\sigma^{-}$ and $\pi$. Thus the the phase difference induced by the differential AC Stark shift is estimated to $4\times10^{-7}\,$Rad$/$s. The phase difference induced by the differential AC Stark shift from $866\,$nm laser is estimated to $2\times10^{-6}\,$Rad$/$s, with $\mathcal{P}=0.05$.

With the above estimations, the differential AC Stark shift is dominated by the operation (1), the picosecond $397\,$nm excitation pulse. The induced phase difference in the memory qubit at $120\,$ms storage time is at the level of $4\times10^{-4}\,\text{Rad}/\text{s}\times120\,\text{ms}=5\times10^{-5}\,$Rad, and the induced infidelity should be on the same level or lower. This infidelity at the level of $10^{-5}$ is much smaller than the $10^{-3}$ statistical error for the measurements demonstrated in Fig.~4. For other storage time of $40\,$ms or $200\,$ms, the influence should also be similar with the $120\,$ms case as the number of noisy operations is linear with time.  Therefore, the measured memory qubit fidelity is dominated by statistical error. In addition, it is also noteworthy that, the differential AC Stark shift does not necessarily lead to decoherence. This shift can be compensated if it is stable. If the effect of the AC Stark difference can be well calibrated and compensated, the resulted decoherence can be further suppressed. This phase difference can also be reduced if the picosecond excitation laser is focused~\cite{lanyon_connection}.

\subsection{Section S14. Heating effects in the ion-photon entangling attempt}

In this section we discuss the heating in the motional state of the two-ion crystal induced by the ion-photon entangling attempts in communication ion. Our trap has trapping frequencies of $(\omega_x, \omega_y, \omega_z)=2\pi\times (1.65, 1.55, 0.5)\,$MHz in three dimensions.

During the ion-photon entanglement generation process, we apply three different operations to the communication qubit, including: (1) a $10$-ps $397\,$nm pulse to excite the ion-photon entanglement in each attempt, (2) an optical pumping with a $\sigma^{+}$-polarization $397\,$nm pulse and a $866\,$nm repumping pulse, both with duration of $290\,$ns before the excitation pulse in each attempt, and (3) $200\,\mu$s sympathetic EIT cooling~\cite{EIT} every $100$ attempts. In these three operations, the first two can induce heating to the ion motional state, and (3) provides the needed cooling to reach equilibrium.

We first estimate the heating due to operation (1) excitation pulse. In each excitation, the communication ion absorbs a $397\,$nm photon with a chance of $95\%$ (the excitation probability is roughly $95\%$ in our experiment). The ion then decays to $|S_{1/2}\rangle$ state with high probability and emits a $397\,$nm photon in a solid angle of $4\pi$ (here we have ignored the small decay probability to $|D_{3/2}\rangle$ with $866\,$nm photon and roughly regard all the emitted photons are in $397\,$nm. This slightly overestimates the heating by a few percents). Thus the expected recoil energy should be $\Delta E=0.95\times((\frac{h}{\lambda_{397\,\text{nm}}})^2/2m)=2\times10^{-29}\,$J. This recoil energy equally heats all the $6$ motional modes of the two-ion crystal, thus $\frac{\Delta E}{6}$ is added to each motional mode. Here we use the radial COM (center-of-mass) mode in $x$ axis $\omega_{rad}^{(x)}=2\pi\times1.65\,$MHz as an example. In each excitation, the added phonon number is $\frac{\Delta E}{6}/\hbar\omega_{rad}^{(x)}=3\times10^{-3}\,$ per attempt.

\begin{figure}
	\centering
	\includegraphics[width=\linewidth]{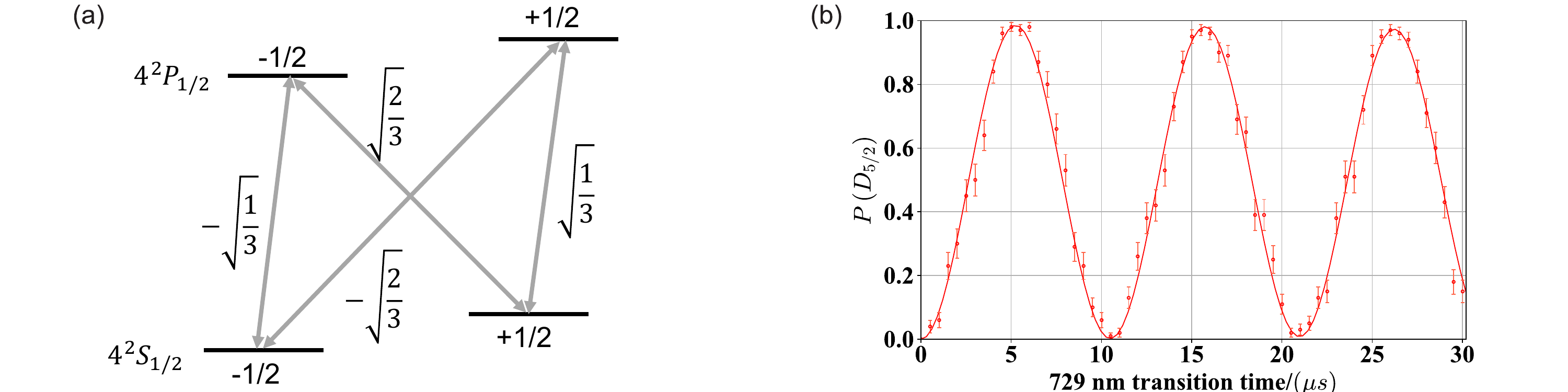}
        \caption{CG coefficient of $397\,$nm transition and $729\,$nm Rabi flop. (a) The CG coefficient of $397\,$nm transition. (b) Calibration of the in-sequence $729\,$nm rotation.}
\end{figure}

For the operation (2) optical pumping, roughly $5$ rounds of excitation to $|P_{1/2},m=+1/2\rangle$ by $\sigma^{+}$ $397\,$nm laser and following spontaneous decay are executed in the $290\,$ns optical pumping to remove the $1/3$ occupation of the unwanted $|S_{1/2},m=-1/2\rangle$ after the picosecond excitation pulse performed in the last attempt (see Fig.~1(b) and Fig.~S12(a)). The unwanted occupation in the $|S_{1/2},m=-1/2\rangle$ state survives with a chance of $2/3$ after each round according to the CG-coefficient, and optical pumping fidelity of $96\%$ to the desired $|S_{1/2},m=+1/2\rangle$ state can be achieved after the optical pumping. Once the ion is in the desired state $|S_{1/2},m=+1/2\rangle$, the optical pumping process stops due to selection rule. Thus the expected number of emitted $397\,$nm photons is $\frac{2}{3}\times0+(\frac{1}{3})^2\times1+(\frac{1}{3})^2\frac{2}{3}\times2+(\frac{1}{3})^2(\frac{2}{3})^2\times3+(\frac{1}{3})^2(\frac{2}{3})^3\times4+\frac{1}{3}(\frac{2}{3})^4\times5=0.87$ per optical pumping. Here we also ignored $866\,$nm photon emission for simplicity, with a few percent difference.

Therefore, the additional phonon number gained in each attempt can be derived by combining the heating effects in both operations (1) and (2) discussed above. Here the total number of phonon added in each attempt is $5.8\times10^{-3}\,$ or $6.2\times10^{-3}\,$ for the two COM modes $\omega_{rad}^{(x)}=2\pi\times1.65\,$MHz or $\omega_{rad}^{(y)}=2\pi\times1.55\,$MHz. The  number of phonon added in each attempt is $6.1\times10^{-3}\,$ or $6.5\times10^{-3}\,$ for the two `rocking' modes at $\omega_{R}^{(x)}=2\pi\times1.57\,$MHz or $\omega_{R}^{(y)}=2\pi\times1.47\,$MHz. Thus one can see that, the expected heating in $100$ attempts is roughly about $6\times10^{-3}\times100=0.6$ phonon in each of the $4$ radial modes of the two-ion crystal. It is noteworthy that as all our laser beams for quantum gates are perpendicular to the axial direction, the motional state of the axial modes will not influence the quality of the quantum gates. We estimate the phonon number of the axial modes are slightly higher than the radial modes.

Here we characterize the effect of the operation (3) sympathetic EIT cooling. In Fig.~S12(b), we demonstrate an in-sequence calibration of our $729\,$nm addressing beam which drives a Rabi flop on memory qubit. The sequence here is very similar with the sequence used in our experiment (Fig.~3(a)) with ion-photon excitation and optical pumping applied to communication qubit each attempt, and the intermediate sympathetic cooling is applied every $100$ attempts. All the parameters of the optical pumping, photon excitation and cooling are the same with the experimental sequence. The memory qubit is shelved to $|D_{5/2},m=+5/2\rangle$ during the operations on communication qubit. This Rabi flop is measured immediately after the sympathetic cooling and an optical pumping is applied to the memory qubit to prepare the memory qubit to $|S_{1/2},m=+1/2\rangle$ before the Rabi flop. The measurement is taken after $1000$ ion-photon entangling attempts (after $10\,$ sympathetic EIT cooling)  to reach the equilibrium in ion crystal temperature. Thus this calibration is under the same condition with the memory qubit in the experimental sequence. We fit the Rabi flop to extract the average phonon number of the $4$ radial modes~\cite{roos thesis}. The result shows that the average phonon number in these $4$ radial modes is $\bar{n}=0.26\pm0.06$. Considering the added phonon number is about $6\times10^{-3}\times100=0.6$ in the following $100$ attempts, then the phonon number ranges from $\bar{n}=0.26$ to $\bar{n}=0.86$ in the experiment. Thus the average phonon number is about $\bar{n}=0.56$ for the radial modes, which can already guarantee high-fidelity operations.

\subsection{Section S15. Success rate in single photon interference}

Here we discuss the success probability of heralded ion-ion entanglement by combining two setups demonstrated in this paper via single photon interference. Here we assume the ion-photon state after the excitation in either trap a or trap b is:
\begin{equation}
|\psi\rangle_{a(b)}=\sqrt{1-\chi}|\downarrow\rangle_{a(b)}|0\rangle_{a(b)}+\sqrt{\chi}|\uparrow\rangle_{a(b)}|1\rangle_{a(b)}
\end{equation}
where $|\downarrow\rangle$ and $|\uparrow\rangle$ are the ion bases. $|0\rangle_{a(b)}$ represents no photon is emitted, and $|1\rangle_{a(b)}=a^{\dag}(b^\dag)|0\rangle_{a(b)}$ represents one photon is emitted from trap a or trap b, as shown in Fig.~S13(b). Here the difference with the ion-photon entanglement in the main text is that, we only collect one polarization in generating ion-photon correlation in the single photon entangling scheme. $\chi\ll 1$ is the excitation probability in each attempt. Here we assume the success probability of photon detection from each trap is $p_a=p_b=p=\chi\eta$ per attempt, where $\eta$ is the total detection efficiency including all the transmission loss in optical channel and detector efficiency. The joint state of both ions and photons is:
\begin{equation}
|\psi\rangle_{a,b}=(1-\chi)|\downarrow\downarrow\rangle_{a,b}|00\rangle_{a,b} + \sqrt{(1-\chi)\chi}(|\downarrow\uparrow\rangle_{a,b}|01\rangle_{a,b}+|\uparrow\downarrow\rangle_{a,b}|10\rangle_{a,b})+\chi|\uparrow\uparrow\rangle_{a,b}|11\rangle_{a,b}
\end{equation}

In the single photon heralded entangling scheme, the two photonic modes $a$ and $b$ from each side are combined in a $50$:$50$ beamsplitter, as shown in Supplementary Fig.~13(b). The two output modes $c$ and $d$ are each sent into a detector. The transformation between the input and output modes in a $50$:$50$ beamsplitter can be expressed as:
\begin{equation}
a^\dagger \rightarrow \frac{c^\dagger+d^\dagger}{\sqrt{2}}  \qquad b^\dagger \rightarrow \frac{c^\dagger-d^\dagger}{\sqrt{2}}
\end{equation}
Combining Eq.~(S5) and Eq.~(S6), the joint state after interfering on the beamsplitter is:
\begin{eqnarray}
|\psi\rangle_{a,b} &=& (1-\chi)|\downarrow\downarrow\rangle_{a,b}|00\rangle_{a,b} +\sqrt{(1-\chi)\chi}(|\downarrow\uparrow\rangle_{a,b}\frac{|10\rangle_{c,d}-|01\rangle_{c,d}}{\sqrt{2}}+|\uparrow\downarrow\rangle_{a,b}\frac{|10\rangle_{c,d}+|01\rangle_{c,d}}{\sqrt{2}})+O(\chi)\\
&=& (1-\chi)|\downarrow\downarrow\rangle_{a,b}|00\rangle_{a,b} +\sqrt{(1-\chi)\chi}(\frac{|\uparrow\downarrow\rangle_{a,b}+|\downarrow\uparrow\rangle_{a,b}}{\sqrt{2}}|10\rangle_{c,d}+\frac{|\uparrow\downarrow\rangle_{a,b}-|\downarrow\uparrow\rangle_{a,b}}{\sqrt{2}}|01\rangle_{c,d})+O(\chi)
\end{eqnarray}
Here we assume $\chi\ll1$ ($\chi=0.02$ in our case, see analysis below) and ignore the double excitation item $O(\chi)$ for simplicity.

\begin{figure}
	\centering
	\includegraphics[width=\linewidth]{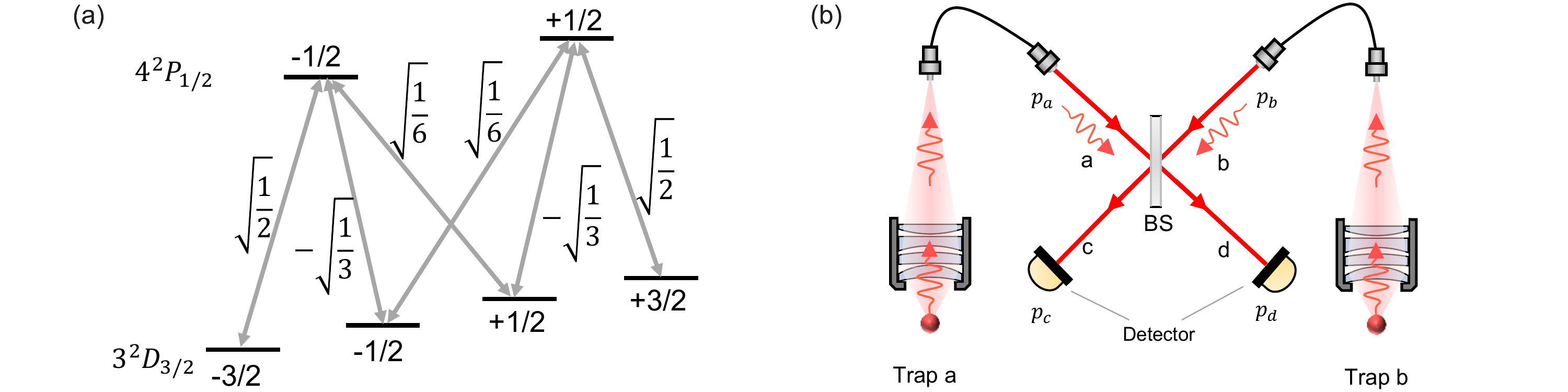}
        \caption{CG coefficient of $866\,$nm transition and $729\,$nm Rabi flop. (a) The CG coefficient of $866\,$nm transition. (b) The scheme of single photon interference.}
\end{figure}

We can clearly see that from Eq.~(S8), the success probability by detecting a photon on detector c (see Fig.~S13(b)) is $p_c=(1-\chi)\chi\eta\approx\chi\eta=p$, as $\chi\ll1$. In this case, the two remote ions a and b are projected to a maximally entangled state $\frac{|\uparrow\downarrow\rangle_{a,b}+|\downarrow\uparrow\rangle_{a,b}}{\sqrt{2}}$. Detector d also clicks with a success probability of $p_d=p$, and the two-ion state is projected to another maximally entangled state $\frac{|\uparrow\downarrow\rangle_{a,b}-|\downarrow\uparrow\rangle_{a,b}}{\sqrt{2}}$. Therefore, no matter which detector clicks, we can produce a remote ion-ion entanglement, and the total success probability by recording the clicks of both detectors is $p_c+p_d=p+p=2p$.

In single photon interference, one can only collect photon with one specific polarization, and all the orthogonal polarization components are discarded, which leads to lower success probability compared to the ion-photon entanglement generation where both orthogonal polarization components can be collected (for example, in our current experiment). In the experiment demonstrated in this paper, we collect three photon transitions which are the $\sigma^{+}$ transition $|D_{3/2},m=-3/2\rangle\rightarrow|P_{1/2},m=-1/2\rangle$, the $\pi$ transition $|D_{3/2},m=-1/2\rangle\rightarrow|P_{1/2},m=-1/2\rangle$ and the $\sigma^{-}$ transition $|D_{3/2},m=+1/2\rangle\rightarrow|P_{1/2},m=-1/2\rangle$, as shown in Eq.~(1) and the following analysis in the main text. These three transitions have a relative weight of $3:4:1$, when collected perpendicular to the magnetic field (CG coefficient is shown in Fig.~13(a), also see Eq.~(2)). This corresponds to an excitation probability $\chi_{\text{ent}}=6\%\times(\frac{1}{3}+\frac{\frac{1}{2}+\frac{1}{6}}{2})=6\%\times\frac{2}{3}=4\%$. In the single photon interference scheme, we only collect the $\pi$ transition $|D_{3/2},m=-1/2\rangle\rightarrow|P_{1/2},m=-1/2\rangle$ (discarding the other two) and achieve a photon excitation probability of $\chi=6\%\times\frac{1}{3}=2\%=\frac{1}{2}\chi_{\text{ent}}$ and a photon generation probability of $p=\chi\eta=0.5\chi_{\text{ent}}\eta=0.5p_{\text{ent}}$, where $p_{\text{ent}}=\chi_{\text{ent}}\eta$ is the success probability of the current experiment. Thus after interference on the beamsplitter, we can have a total success probability of $2p=2\times0.5p_{\text{ent}}=p_{\text{ent}}$ to generate a remote ion-ion entanglement (see discussion above), which has the same success probability with the ion-photon entanglement generation in the current experiment.

\subsection{Section S16. Modulation in memory qubit fidelity and the limitation to coherence time}

In Fig.~4(a) the memory qubit fidelity is modulated with two frequencies at $50\,$Hz and $150\,$Hz. The $50\,$Hz and $150\,$Hz modulation mainly comes from the power line of the AC electricity. The AC power in our lab has a frequency of $50\,$Hz. In the experiment we have already enclosed our setup into a double-layer Mu-metal box. The Mu-metal box has a size of $1\,\text{m}\times1\,\text{m}\times1\,\text{m}$, and we can place our chamber and the nearby optics into the box. Each layer of the Mu-metal has a thickness of $2\,$mm  with an aluminum board of $4\,$mm thick in the middle. Although a magnetic field shielding of $17.38\,$dB and $11.11\,$dB at $50\,$Hz and $150\,$Hz (third order harmonics of $50\,$Hz) have been achieved, the strong magnetic field components at $50\,$Hz and $150\,$Hz from the power line still dominate the modulation shown in Fig.~4(a). It is worth to note that these modulations at $50\,$Hz and $150\,$Hz are in fact coherent, and in principle the added phase can be compensated according to storage time. Thus they are not the limiting factors of the $323\,$ms decoherence time in our experiment.

We attribute the main limiting source of the memory coherence time $323\,$ms to the remaining magnetic field noise in other frequencies (except power line frequencies $50\,$Hz and $150\,$Hz). Although the $50\,$Hz and its harmonics are the strongest magnetic field components in our lab, there exist other frequency components which induce the memory decoherence. These frequency components are already shielded by our Mu-metal box but still contribute to the memory decoherence on a time scale of $100\,$ms, even with the spin echo in our experiment. In addition, the memory qubit is encoded in the superposition of $|D_{5/2},m=+3/2\rangle$ and $|D_{5/2},m=+5/2\rangle$. These two states have very different energy shifts according to magnetic field ($g_j\mu_B\Delta m\sim1.68\,$MHz$/$Gauss), thus the qubit stored as a superposition of these two states is quite sensitive to the magnetic field noise in the lab.

\subsection{Section S17. Comparison with dual-species scheme}

In this section, we compare the dual-type scheme used in our experiment with the dual-species scheme (\cite{oxford distributed,oxford_sr_and_ca}). Specifically, we will compare the crosstalk effects in both the internal state and the motional state under these two schemes, as well as the performance in applications such as distributed quantum computing.

For both schemes, the crosstalk effects on the internal states are mainly determined by: 1) the spectral separation between the optical transitions for communication qubit and memory qubit, and 2) the light intensity at the position of memory qubit during the operations on communication qubit, as discussed in Section S13. Specifically, for the dual-type scheme in our experiment, the $397\,$nm and $866\,$nm transitions used for the communication qubit are about $400\,$THz and $5\,$THz apart from the $854\,$nm transition which affects the memory qubit (the weak $729\,$nm transition can be ignored here). For the dual-species scheme (\cite{oxford distributed,oxford_sr_and_ca}), the spectral separations are about $50\,$THz between the $422\,$nm transition for Sr$^+$ communication qubit and the closest $397\,$nm or $393\,$ nm transitions for Ca$^+$ memory qubit. Although the $866\,$nm repumping light has a relatively small $5\,$THz spectral separation in our case (about $10$ times smaller than the $50\,$THz in the dual-species case), the crosstalk influence induced by the $866\,$nm light is not that large, as the intensity of the $866\,$nm repumping beam is usually much lower than the strong UV light ($422\,$nm in dual-species case or $397\,$nm in our case) in cooling, pumping, and ion-photon excitation. Except for spectral separation and optical intensity mentioned above, the crosstalk effects are also influenced by many other factors, which can be quite complicated in the real case. For example, the crosstalk effects are also influenced by the different polarizations of the lasers, as described in Section S13. The energy level structures in these two cases are also different. There are two strong UV transitions ($397\,$nm and $393\,$nm) of the memory qubit which can be affected by the strong $422\,$nm light in the dual-species scheme, compared to only one $854\,$nm transition which can be influenced by $397\,$ nm and $866\,$nm lasers in our case. However, despite all these differences, the final crosstalk effect in these two schemes are in fact very similar. The measured light shift in~\cite{oxford_sr_and_ca} is $8.8\times10^{-6}$ Rad/attempt, which is very close to the value of $8\times10^{-6}$ Rad/attempt we estimated in Section S13 for our case.

Both schemes also have very similar performances regarding the crosstalk effect in motional state. The measured heating in each ion-photon entangling attempt is about $9.3\times10^{-4}$ phonon/attempt in the dual-species scheme (\cite{oxford_sr_and_ca}), which is in the same order of magnitude as the estimated heating of $3\times10^{-3}$ phonon/attempt in our case (see Section S14). The difference can be explained by the different mass of the communication qubit ($^{88}$Sr$^+$ in \cite{oxford distributed,oxford_sr_and_ca} and $^{40}$Ca$^+$ in our case) as heavier ion picks up less recoil energy in each attempt ($\Delta E\sim1/m$, see Section S14) and the different trap frequencies in these two cases. Meanwhile, it is also noteworthy that the dual-type scheme has an advantage over dual-species in sympathetic cooling. The ratio of mass between the communication qubit and the memory qubit in the dual-type scheme is exactly $1$, which is the optimal value for sympathetic cooling (see \cite{ratio}). The mass ratio in \cite{oxford distributed,oxford_sr_and_ca} is about $88/43\approx2$, which is still good but not as optimal as the case in the dual-type scheme. Therefore, the dual-type scheme could be more advantageous over the dual-species scheme in removing the motional excitations of the ion crystal via sympathetic cooling. In \cite{oxford_sr_and_ca}, the measured phonon numbers are $0.3$ and $3$, which are similar to our case of $0.26$ to $0.86$ estimated in Section S14.

Therefore, both the dual-type scheme and the dual-species scheme can guarantee a negligible crosstalk in internal state and maintain a low motional excitation number simultaneously, as results of the large spectral separation over THz between the communication qubit and memory qubit and the removal of motional excitations via sympathetic cooling.

Here we also compare how distributed quantum computing performs in the cases when the memory coherence time ($\tau_D$) and the time for remote entanglement generation ($\tau_E$) are of a similar scale and when $\tau_D$ is much longer than $\tau_E$.

The influence of the ratio $\tau_E/\tau_D$ on the distributed quantum computing has been extensively studied in previous literature (\cite{duan pra}). It is shown in \cite{duan pra} that scalable fault-tolerant distributed quantum computing can be realized for any ratio $\tau_E/\tau_D$, but at the cost of increased overhead in qubits and time when $\tau_E/\tau_D$ is large. The comparison between the cases when the ratio $\tau_E/\tau_D$ is close to unity (our case) and the case when $\tau_E/\tau_D\ll1$ (\cite{oxford distributed,oxford_sr_and_ca}) is clearly illustrated in Fig.~8(c) of \cite{duan pra}. One can see that with a smaller $\tau_E/\tau_D$, fewer resources are needed to support distributed quantum computing. The method to achieve fault-tolerant quantum computing is also demonstrated in detail in \cite{duan pra}.

Although scalable fault-tolerant distributed quantum computing is possible for any ratio $\tau_E/\tau_D$, it is still advantageous and important to achieve a ratio $\tau_E/\tau_D\ll1$ to make the experimental realization practical. The small $\tau_E/\tau_D$ demonstrated in previous works with dual-species scheme (\cite{oxford distributed,oxford_sr_and_ca}) can significantly save the computing resources according to \cite{duan pra}. In our case, although the current ratio $\tau_E/\tau_D$ is about $0.1$ in the $3\,$m fiber case, we can reduce this ratio to below $10^{-2}$ with future improvements discussed in Section S8 and Table S1. This ratio can be further improved if cavity enhancement is employed in future. With these improvements, our scheme will also be able to implement distributed quantum computing with high performances in the future.

\end{document}